\documentclass[%
 reprint,
superscriptaddress,
preprintnumbers,
 amsmath,amssymb,
 aps,
prb,
]{revtex4-1}

\usepackage{subfigure}
\usepackage{graphicx}
\usepackage{dcolumn}
\usepackage{bm}

\usepackage{hyperref}
\hypersetup{colorlinks=true,allcolors=blue}

\begin{document}
\title{Transport in Conductors and Rectifiers: Mean-Field Redfield Equations and Non-Equilibrium Green's Functions}
\author{Zekun Zhuang}
\affiliation{Department of Physics, Box 1843, Brown University, Providence, Rhode Island 02912-1843, USA}
\author{Jaime Merino}
\affiliation{Departamento de Fs\'{i}ca Te\'{o}rica de la Materia Condensada, Condensed Matter Physics Center (IFIMAC) and Instituto Nicol\'{a}s
Cabrera, Universidad Aut\'{o}noma de Madrid, Madrid 28049, Spain
}
\author{J.B. Marston}
\affiliation{Brown Theoretical Physics Center and Department of Physics, Box 1843, Brown University, Providence, Rhode Island 02912-1843, USA}
\date{\today }

\begin{abstract}
We derive a closed equation of motion for the one particle density matrix of a quantum system coupled to multiple baths using the Redfield master equation combined with a mean-field approximation. The steady-state solution may be found analytically with perturbation theory.  Application of the method to a one-dimensional non-interacting quantum wire yields an expression for the current that reproduces the celebrated Landauer's formula.  Nonlinear rectification is found for the case of a mesoscopic three-dimensional semiconductor p-n junction. The results are in good agreement with numerical simulations obtained using non-equilibrium Green's functions, supporting the validity of the Redfield equations for the description of transport.  
\end{abstract}

\maketitle


\section{Introduction}
Open quantum systems present an interesting challenge for theory. One popular approach relies on  a quantum master equation (QME) and the Born-Markov approximation \cite{redfield1957,lindblad1976,breuer2002,ishizaki2009,rivas2012,purkayastha2017}. QMEs can be used to investigate the transport properties in condensed matter systems, for example in spin chains or fermionic or bosonic tight-binding models.  The Lindblad equations, a specific kind of Markovian QME, have been frequently used   \citep{wichterich2007,prosen2008,benenti2009,prosen2011open,prosen2011exact,vznidarivc2013,santos2016,guimaraes2016,reichental2018}. The Lindblad equations maintain positivity of the density matrix (all eigenvalues are non-negative), but there are difficulties with the definition of the currents and with thermalization \citep{gebauer2004,bodor2006,wichterich2007,ishizaki2009,salmilehto2012,purkayastha2016, fleming2011,guimaraes2016,santos2016,Hovhannisyan2019,kirvsanskas2018,reichental2018}. The Redfield equation is an alternative QME that, despite not respecting the positivity of density matrix for some initial conditions \cite{gnutzmann1996,palmieri2009,farina2019}, may describe the dynamics more accurately \citep{purkayastha2016} since it does not require further approximation beyond the  Born-Markov approximation such as the secular (or rotating-wave) approximation. The secular approximation is frequently invoked in derivations of the Lindblad equations but can be problematic under some circumstances \cite{fleming2010} and can break down in the limit of large system size, especially for gapless systems. A separate problem for interacting systems is the exponential growth of the many-body Hilbert space and the dimension of the density matrix with system size \citep{xu2019,Nagy2019,Hartmann2019}. In this paper we treat interactions in a mean-field approximation to avoid this problem.  

A parallel approach to QMEs employs non-equilibrium Green's functions (NEGF) \cite{Schwinger1961,Kadanoff1962,Keldysh1964} that extend the standard equilibrium formalism \cite{Abrikosov,Fetter,Mahan} 
and provides, in principle, a systematic approach to deal with out-of-equilibrium quantum many-body systems. The NEGF approach has been successfully applied to a broad variety of phenomena ranging from 
electronic transport through semiconductors \cite{Jauho} and nanostructures  \cite{Yigal1992} to ultrafast pump-probe spectroscopies of strongly correlated materials. The combination of NEGF with dynamical mean-field theory (DMFT) provides \cite{Aoki2014} a 
powerful tool to describe, for instance, electronic currents through strongly correlated materials or the time-evolution of cold atoms in optical lattices under sudden quenches. 
The NEGF approach is formally exact.  In principle it avoids the Born-Markov approximations.  The Kadanoff-Baym equations of motion \cite{Kadanoff1962} describing the 
time-evolution of an open quantum many-body system are causal and make no assumptions on the strength of the interactions in the system nor on the strength of the coupling of the system to the environment. The equivalent Keldysh formulation of NEGF 
provides a precise description of the steady state of an open quantum system in which the initial electronic correlations are lost. An exact expression for the stationary current through a system coupled 
to large fermionic leads can be obtained \cite{Yigal1992} making it useful for benchmarking less computationally intensive methods such as the Redfield approach discussed below.  

Here we examine the validity of the Redfield QMEs for the study of non-equilibrium properties, especially transport, of weakly-interacting systems.  To do this we formulate a modified Redfield equation (MRE) by dropping the imaginary Cauchy principal value parts (CPVP) of the dissipator.  Within a mean-field approximation such MREs yield a closed equation of motion for the one-particle density-matrix (OPDM) which can be solved. To test the MRE approach we apply it to a one-dimensional quantum wire and to a three-dimensional semiconductor p-n junction, investigate transport analytically and numerically, and show that the results agree well with NEGF methods.

The rest of the paper is organized as follows. A brief introduction to the full Redfield equations is presented in Section \ref{FullRedSec} with details left to Appendix  \ref{AppRedfieldDerivation}. Section \ref{SecModified} discusses the MRE and the mean-field approximation is made in Section \ref{SecMF}. Section \ref{SecNum} briefly discusses the numerical method used to find the self-consistent solution of MRE and Section \ref{Connection} discusses the connection between Redfield equations and Lindblad equations.  Section \ref{SecNEGF} introduces the NEGF method we use. Section \ref{SecApp} applies the MRE method to different systems, and includes a comparison to results obtained by NEGF.  Conclusions are discussed in Section \ref{Conclusion}.

\section{The Redfield equations}

\subsection{Full Redfield equation}\label{FullRedSec}
\begin{figure}
\centering
\includegraphics[width= 3 in]{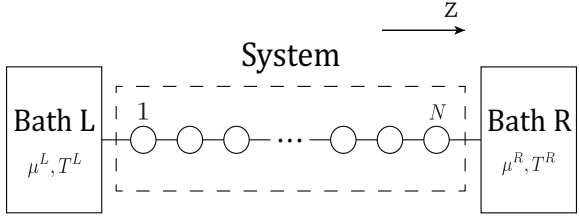}
\caption{Schematic of the total system that is made up of the 1-D tight-binding chain and baths (or leads) at each end.}
\label{Fig1}
\end{figure}

In this section we derive the full Redfield equations. For concreteness we consider the specific problem of a one-dimensional tight-binding chain, connected to separate baths (leads) at each end as shown in Fig. \ref{Fig1}. 
The total Hamiltonian including the chain and the baths is:
\begin{equation}
H_{\text{total}}=H_S+H_B^{(L)}+H_B^{(R)}+H_{SB}^{(L)}+H_{SB}^{(R)},
\end{equation} 
where $H_S = H_S^0 + V_S$, 
\begin{equation}
H_S^0=-t_S\sum_{i=1}^{N}\sum_{\sigma=\uparrow,\downarrow}(c_{i,\sigma}^{\dagger}c_{i+1,\sigma}+\text{H.c.})+\sum_{i=1}^{N} \mu_i n_i  
\label{sysH}
\end{equation}
is the non-interacting system Hamiltonian,
\begin{equation}
V_S = \frac{1}{2}\sum_{i,j,\sigma,\sigma^\prime}V_{i,\sigma;j,\sigma^\prime}n_{i,\sigma}n_{j,\sigma^\prime}
\end{equation}
is the interaction, 
\begin{equation}
H_B^{(\alpha)}=\sum_{\lambda,\sigma}\epsilon_{\lambda,\sigma}^{\alpha}f^{\alpha \dagger}_{\lambda,\sigma}f_{\lambda,\sigma}^\alpha, \quad \alpha=L\  \text{or}\  R
\end{equation}
is the Hamiltonian for the left and right baths, and
\begin{equation}
H_{SB}^{(\alpha)}=\sum_{\lambda,\sigma}(T_{\lambda}^\alpha f_{\lambda,\sigma}^{\alpha \dagger} c_{i_\alpha,\sigma}+T_{\lambda}^{\alpha *} c_{i_\alpha,\sigma}^\dagger f_{\lambda,\sigma}^\alpha), \quad \alpha=L\  \text{or}\  R \label{coupling}
\end{equation}
models the coupling between the system and the baths.  
Operator $c_{i,\sigma}^\dagger$ $(c_{i,\sigma})$ creates (annihilates) a fermion at chain site $i$ with spin $\sigma$, operator $f_{\lambda,\sigma}^{\alpha \dagger}$  ($f_{\lambda,\sigma}^\alpha$) creates (annihilates) a spin $\sigma$ fermion in a bath $\alpha$ eigenstate $\lambda$ with eigenenergy $\epsilon^\alpha_{\lambda,\sigma}$, $\mu_i$ is the on-site potential and $V_{i,\sigma;j,\sigma^\prime}$ is the interaction matrix element. We have set $i_\alpha=1$ if $\alpha=L$ and $i_\alpha=N$ if $\alpha=R$. In the following for simplicity we suppress the spin index $\sigma$ unless specified otherwise.

Treating the combination of the two baths and the chain as a whole, the equation of motion (EOM) for the density matrix is given by:
\begin{equation}
\frac{\partial \rho_{\text{total}}}{\partial t}=-i[H_{\text{total}},\rho_{\text{total}}]
\end{equation}
where we have set $\hbar=1$ and in the following. 

To obtain the EOM for the reduced density matrix of the system $\rho_S=\text{Tr}_B\rho_{\text{total}}$, the usual approach is to make the Born-Markov approximation to trace out the degrees of freedom of the baths \cite{breuer2002,rivas2012}. The approximation assumes that the coupling between system and baths is weak, that the baths are too large to be affected by the system, and that the baths retain no memory of past history. A detailed derivation and discussion may be found in Appendix \ref{AppRedfieldDerivation} and the final result is
\begin{equation}
\frac{d\rho_S}{dt}=-i[H_S,\rho_S] + L(\rho_S) \label{FullRedfield1}
\end{equation}
where $L(\rho_S)$ is called the ``dissipator'':
\begin{multline}
L(\rho_S) = -\sum_{\alpha,\Omega,\sigma}\left\lbrace (F^\alpha(\Omega)+iF^{\alpha I}(\Omega))\times[c_{i_\alpha},[c_{i_\alpha}(\Omega)]^\dagger \rho_S]\right.\\+\left.(H^\alpha(\Omega)-iH^{\alpha I}(\Omega))\times[c_{i_\alpha}^\dagger,c_{i_\alpha}(\Omega)\rho_S]+\text{H.c.} \right\rbrace\ . \label{FullRedfield2}
\end{multline}
Here $F^\alpha(\Omega)=\pi N^\alpha(\Omega)f^\alpha(\Omega)$, $F^{\alpha I}(\Omega)=\frac{1}{\pi}\text{P}\int_{-\infty}^{\infty}d\omega\frac{F^\alpha(\omega)}{\omega-\Omega}$, $H^\alpha(\Omega)=\pi N^\alpha(\Omega)(1-f^\alpha(\Omega))$, $H^{\alpha I}(\Omega)=\frac{1}{\pi}\text{P}\int_{-\infty}^{\infty}d\omega\frac{H^\alpha(\omega)}{\omega-\Omega}$. $N_\sigma^\alpha(\Omega)=\sum_\lambda \vert T_\lambda^\alpha\vert^2\delta(\Omega-\epsilon^\alpha_{\lambda,\sigma})$ is the density of states for bath $\alpha$, and $f^\alpha(\Omega)=1/(e^{(\Omega-\mu^\alpha)/k_B T^\alpha}+1)$ is the Fermi-Dirac distribution function. $T^\alpha$ and $\mu^\alpha$ are the temperature and the chemical potential of the bath $\alpha$ respectively. The eigenoperators $A(\Omega)$ are defined by:
\begin{equation}
A(\Omega)=\sum_{\epsilon^\prime-\epsilon=\Omega}\Pi(\epsilon)A\Pi(\epsilon^\prime) \label{eigOp}
\end{equation}
where $\Pi(\epsilon)=\sum_i\vert\epsilon_i\rangle\langle\epsilon_i \vert$ is the projection operator onto the subspace spanned by all the eigenstates with energy $\epsilon$. When the system is non-interacting, $[c_i(\Omega)]^\dagger=\sum_\lambda\langle \lambda\vert i\rangle c_\lambda^\dagger \delta_{\epsilon_\lambda,\Omega}$ and $c_i(\Omega)=\sum_\lambda\langle i\vert \lambda\rangle c_\lambda \delta_{\epsilon_\lambda,\Omega}$, where $\vert\lambda\rangle $ are the system eigenstates. 

Though Eq. (\ref{FullRedfield1}) and Eq. (\ref{FullRedfield2}) are derived without further approximation beyond Born-Markov, it is well known that the Redfield equation may violate the positivity of the density matrix in some circumstances and does not necessarily give the Gibbs distribution  at equilibrium. Nevertheless, dropping the imaginary CPVP in Eq. (\ref{FullRedfield2}) yields a modified Redfield equation that does give the desired Gibbs distribution at equilibrium. To see this first note that
\begin{equation}
[H_S,A(\Omega)]=-\Omega A(\Omega)\ .
\end{equation}
Application of the Baker-Campbell-Hausdorff identity yields
\begin{equation}
e^{\beta H_S}[c_{i}(\Omega)]^\dagger e^{-\beta H_S}=e^{\beta \Omega}[c_{i}(\Omega)]^\dagger\label{EigOpRelation1}
\end{equation}
and
\begin{equation}
e^{\beta H_S}c_{i}(\Omega)e^{-\beta H_S}=e^{-\beta \Omega}c_{i}(\Omega)\ .
\label{EigOpRelation2}
\end{equation}
Substituting the equilibrium thermal distribution $\rho_{th}=e^{-\beta H_S}/\text{Tr}(e^{-\beta H_S})$ into Eq. (\ref{FullRedfield2}), combined with Eq. (\ref{EigOpRelation1}) and Eq. (\ref{EigOpRelation2}), and dropping all CPVPs, shows that $L(\rho_{th})=0$ if the bath chemical potential is set to zero, which means the left hand side of the Eq. (\ref{FullRedfield1}) also vanishes because $H_S$ commutes with $\rho_{th}$. The recovery of the equilibrium thermal distribution thus motivates discarding the CPVP that appears in Eq. (\ref{FullRedfield2}) as discussed in more detail in Section \ref{SecModified}.

\subsection{Modified Redfield equation for the OPDM} \label{SecModified}
Single-particle operators that are bilinear in the fermion creation and annihilation operators are often of greatest interest, and for such observables the full density matrix using Eq. (\ref{FullRedfield1}) and Eq. (\ref{FullRedfield2}) is not needed. Therefore, in the following we focus on the OPDM defined as
\begin{equation}
\rho_{ij}^{(1)}=\langle c_j^\dagger c_i\rangle=\text{Tr}(c_j^\dagger c_i\rho).
\end{equation}
For a non-interacting system coupled to the bath by a single fermion operator as in Eq. (\ref{coupling}), the EOM for the OPDM can be closed. Multiplying both sides of Eq. (\ref{FullRedfield1}) by $c_{j}^\dagger c_{i}$ and taking the trace yields:
\begin{multline}
\frac{d\rho^{(1)}}{dt}=-i[h,\rho^{(1)}]+\sum_\alpha \left\lbrace(\rho_\alpha^{(1)}-\rho^{(1)})\hat{J}^\alpha P_{i_\alpha}\right. \\+\left. iP_{i_\alpha}(\hat{K}^\alpha\rho^{(1)}-\hat{F}^{\alpha I}) +\text{H.c.}\right\rbrace  \label{ODDMRedfield1}
\end{multline}
where $h=\sum_\lambda \vert \lambda\rangle \epsilon_\lambda \langle \lambda\vert$ is the system Hamiltonian in first-quantized form, $\rho_{\alpha}^{(1)}=\sum_\lambda \vert \lambda\rangle f^\alpha(\epsilon_\lambda) \langle \lambda\vert$ is the OPDM of the system with the temperature and chemical potential of bath $\alpha$, $P_{i_\alpha}=\vert i_\alpha\rangle \langle i_\alpha\vert$ is the projection operator that projects onto site $i_\alpha$, and $\hat{J}^\alpha=\pi\sum_\lambda \vert \lambda\rangle  N^\alpha(\epsilon_\lambda) \langle \lambda\vert$ is the matrix related to the spectral density of bath $\alpha$. 
$\hat{F}^{\alpha I}=\sum_\lambda \vert \lambda\rangle F^{\alpha I}(\epsilon_\lambda) \langle \lambda\vert$ and $\hat{K}^\alpha=\sum_\lambda \vert \lambda\rangle \text{P}\int_{\infty}^{\infty} d\omega \frac{N^\alpha(\omega)}{\omega-\epsilon_\lambda} \langle \lambda\vert$ are two matrices arising from the imaginary CPVP in Eq. (\ref{FullRedfield2}). In the wide-band limit $\pi N^\alpha(\epsilon_\lambda)=J^\alpha$, and  $\hat{K}^\alpha$ is nearly proportional to the identity matrix. The additional Lamb shift Hamiltonian, which is often neglected, is given by $H_{LS}=-\sum_{\alpha}P_{i_\alpha}\hat{K}^\alpha$. The Lamb shift term has its origin in the interaction between system and baths, analogous to the energy shift of electrons due to their  interaction with the quantized photon ``bath'' of quantum electrodynamics.  Further neglecting the $[P_{i_\alpha},F^{\alpha I}]$ term yields a much simpler equation for the OPDM:
\begin{equation}
\frac{d\rho^{(1)}}{dt}=-i[h,\rho^{(1)}]+\sum_{\alpha}J^\alpha\{\rho_{\alpha}^{ (1)}-\rho^{(1)},P_{i_\alpha}\}\ .
\label{ODDMRedfield2}
\end{equation}
It is then obvious that the OPDM of the steady state given by Eq. (\ref{ODDMRedfield2}) when the system is only connected to one bath $\alpha$ is just $\rho_{\alpha}^{(1)}$, which implies that Eq. (\ref{ODDMRedfield2}) gives correct Gibbs state at equilibrium, consistent with the analysis of Section \ref{FullRedSec}. Equation (\ref{ODDMRedfield2}) is the central result that will be used in this paper. The generalization to the three-dimensional case with translational invariance in two of the three dimensions is straightforward and only slight modifications are required. For more details and further discussion see Appendix \ref{AppRed3D}. 

We now discuss the decision to drop the CPVPs in Eq. (\ref{FullRedfield2}) and Eq. (\ref{ODDMRedfield1}). Ref. \cite{mori2008} observes that the CPVPs come from the renormalization of the system Hamiltonian.  By ignoring the CPVPs, the steady state given by Redfield equation due to single bath is the desired Gibbs distribution.  In the context of the Lindblad equations, the CPVPs nearly cancel out with the remainder being the Lamb shift  (see Section \ref{Connection}) that is often ignored. 

We now consider  the existence and uniqueness of the steady state solution of Eq. (\ref{ODDMRedfield2}). For the steady state, from Eq. (\ref{ODDMRedfield2}) we obtain
\begin{equation}
A\rho_{st}^{(1)}+\rho_{st}^{(1)}A^\dagger+\sum_\alpha J^\alpha\{\rho^{(1)}_{\alpha},P_{i_\alpha}\}=0 \label{Lyapunov}
\end{equation}
with $A=-ih-\sum_\alpha J^\alpha P_{i_\alpha}$.  This equation is of the Lyapunov, or more generally, Sylvester form.  For a unique solution to exist, $A$ and $-A^\dagger$, or $h-i\sum_\alpha J^\alpha P_{i_\alpha}$ and $h+i\sum_\alpha J^\alpha P_{i_\alpha}$, are forbidden to have common eigenvalues \cite{datta2004}.  This is equivalent to requiring $u_i \neq u_j^*$ $\forall i,j$, where $u_i$ is the $i$-th eigenvalue of $h-i\sum_\alpha J^\alpha P_{i_\alpha}$. The condition is violated when the system has disconnected parts because then the Hamiltonian $h$ can be block diagonalized in the subsystem basis. Numerical solutions of Eq. (\ref{Lyapunov}) also become challenging if $u_i\approx u_j^*$ for some $i,j$ despite the formal existence of a unique solution.

In some instances it is possible to treat $J^\alpha=J$ in Eq. (\ref{Lyapunov}) as a small parameter and obtain a perturbative solution. The solution should be valid for small systems with large spacing between energy levels such as semiconducting quantum dots; see Appendix \ref{AppPerturbation}. Insight into the structure of MRE comes from the lowest-order solution to the OPDM in eigenenergy basis:
\begin{equation}
(\rho^0_\text{{diag}})_{\mu\mu}=\frac{\sum_\alpha (\rho_\alpha)_{\mu \mu} (P_{i_\alpha})_{\mu \mu} }{\sum_\alpha (P_{i_\alpha})_{\mu \mu}}\ . \label{DiagPert}
\end{equation}
This equation says that the density matrix of the system is simply a superposition of the density matrices set by the different baths.

\subsection{Mean-field approximation} 
\label{SecMF}
For a general system with interaction, it is not possible to obtain closed EOM for the OPDM, the quantum analogue of the classical Bogoliubov-Born-Green-Kirkwood-Yvon (BBGKY) hierarchy problem. Following the approach taken in  Ref. \citep{rosati2014,wu2010}, our strategy here is to treat the system Hamiltonian in a mean-field approximation, as then the system is a collection of independent particles with some self-consistently determined parameters.  The full Hamiltonian $h$ is replaced by the mean-field Hamiltonian $h_{MF}$, and $\rho_\alpha^{(1)}$ becomes $\rho_{\alpha,MF}^{(1)}=\sum_{\lambda_{MF}}\vert \lambda_{MF}\rangle f^\alpha(\epsilon_{\lambda}^{MF})\langle \lambda_{MF}\vert$, where $\vert \lambda_{MF}\rangle$ and $\epsilon_\lambda^{MF}$ are the eigenstates and eigenenergies of the mean-field Hamiltonian $H_{MF}$ at each instant time.  The parameters of the mean-field Hamiltonian implicitly depend on $\rho_\alpha^{(1)}(t)$.  The mean-field approximation can also be made from the outset prior to the derivation of the Redfield equation.  As the Markov approximation assumes that $\rho(t)$ does not 
change appreciably over the memory time scale of the baths, the same final result is obtained.

\subsection{Numerical technique}\label{SecNum}
We briefly discuss the numerical technique used to find the non-equilibrium steady-state of the MRE within the Hartree mean-field approximation. The calculation is carried out in the real-space position basis.  First, an initial choice for the mean-field Hamiltonian $h_\text{MF}$ is made. Next $h_\text{MF}$ is used to obtain $\rho^{(1)}_{\alpha}$ as described in Section \ref{SecMF} and the corresponding initial OPDM $\rho^{(1)}_{1}$ is found with the use of  Eq. (\ref{Lyapunov}). The calculation is carried out with a Lyapunov equation solver. The OPDM and the mean-field Hamiltonian are then repeatedly updated using $\rho^{(1)}_{n} = (1-\epsilon) \rho^{(1)}_{n-1} + \epsilon \tilde{\rho}^{(1)}_{n}$ where $0 < \epsilon \ll 1$ is a small parameter that controls the rate of updating, and $\tilde{\rho}^{(1)}_{n}$ is the solution to the Lyapunov equation
\begin{equation}
A \tilde{\rho}^{(1)}_n + \tilde{\rho}^{(1)}_n A^\dagger + \sum_\alpha J^\alpha\{\rho^{(1)}_{\alpha},P_{i_\alpha}\} = 0
\end{equation}
with operator $A$ determined by the OPDM from the previous step, $\rho^{(1)}_{n-1}$.
The procedure continues until the change in the OPDM and the mean-field Hamiltonian is smaller than a predetermined threshold; at that point a good approximation to the self-consistent solution $\rho^{(1)}_{st}$ is at hand. One possible obstruction can occur when, as already noted in Section \ref{SecModified}, $u_i\approx u_j^*$ at an intermediate step.  Such a situation may arise if the on-site potential in space varies too much, and can be avoided by careful choice of the initial Hamiltonian and $\epsilon$. 

\subsection{Connection with Lindblad equations}\label{Connection}
In this section we briefly discuss the connection between the Redfield equations and the more widely used Lindblad equations. The Lindblad dissipator is given by: \cite{breuer2002}
\begin{equation}
L(\rho_S)=\sum_k \left(L_k \rho_S L_k^\dagger-\frac{1}{2}\{L_k^\dagger L_k ,\rho_S\}\right)\ ,
\end{equation}
where $L_k$ is called the Lindblad operator. Depending on whether or not the Lindblad operators are local in space, the Lindblad equations are sometimes classified as local Lindblad equations \citep{vznidarivc2013,benenti2009,prosen2008,prosen2011open,prosen2011exact} or global Lindblad equations \citep{guimaraes2016,santos2016}, and their validity in different circumstances is still under investigation \citep{manrique2015,hofer2017}. 

To derive the global Lindblad equation with Lindblad operator $L\sim c^\dagger_{\lambda}$ or $c_{\lambda}$ typically requires the use of the so-called secular (or rotating-wave) approximation  and omits all the terms with $\Omega\neq \Omega^\prime$ in Eq. (\ref{AppRedfield}) because $e^{i(\Omega-\Omega^\prime)t}$ oscillates rapidly in the interaction picture and average to zero at the coarse-grained time scale $t_S\gg 1/\text{min}\{\vert \Omega-\Omega^\prime\vert\neq 0\}$, where $\text{min}\{\vert \Omega-\Omega^\prime\vert\neq 0\}$ is the smallest nonzero value. However, the treatment inevitably becomes problematic in the thermodynamic limit of large system size, for example in a metal, as the energy level spacing is infinitesimally small and time-scale separation breaks down. Such global Lindblad equations also raise some difficulty in defining the current and it is necessary to include a fictitious current between partitions that are not directly coupled to ensure that the current obeys continuity. To derive the global Lindblad equation from the Redfield equations, Eqs.(\ref{FullRedfield1}) and (\ref{FullRedfield2}), operators $c_{i_\alpha}$ are replaced by $c_{i_\alpha}(-\Omega)$ and $c_{i_\alpha}^\dagger$ by $c_{i_\alpha}^\dagger(\Omega)$ in Eq. (\ref{FullRedfield2}). Consequently, the projection operators $P_{i_\alpha}$ in Eq. (\ref{ODDMRedfield1}) are also replaced by $\tilde{P}_{i_\alpha}=\sum_\lambda \vert \lambda \rangle \vert\langle i_\alpha \vert \lambda \rangle\vert^2 \langle \lambda \vert$ and thus all the off-diagonal elements of $P_{i_\alpha}$ in the energy eigenbasis are ignored. The substitution leads to the decoupling of the diagonal and off-diagonal elements of the OPDM in energy eigenbasis, and since $[\tilde{P}_{i_\alpha},\tilde{F}^{\alpha I}]=0$, the standard global Lindblad equation in terms of OPDM is obtained:
\begin{equation}
\frac{d\rho^{(1)}}{dt}=-i[h+h_{LS},\rho^{(1)}]+\sum_\alpha {J}^\alpha\{(\rho_\alpha^{(1)}-\rho^{(1)}), \tilde{P}_{i_\alpha}\}
\end{equation}
where $h_{LS}=-\sum_\alpha \tilde{P}_{i_\alpha} K^\alpha$ is again the Lamb shift Hamiltonian, due to the interaction between system and reservoirs.

The local Lindblad equation with Lindblad operator $L\sim c^\dagger_{i_\alpha}$ or $c_{i_\alpha}$ can be arrived at by an alternative approximation.  For the chain, if the hopping between sites is weak compared to the memory time of the bath $\tau_B$, namely $t_S \tau_{B}\ll 1$, to a good approximation the local operators evolve solely due to the local Hamiltonian, namely $U^\dagger (t^\prime-t)c_{i_\alpha}^\dagger (0) U(t^\prime-t)\approx e^{-i \mu_{i_\alpha} (t^\prime-t)} c_{i_\alpha}^\dagger$. Inserting this equation into Eqs.  (\ref{C+}) and (\ref{C-}) after transforming back to the Schr\"odinger picture, or alternatively replacing $c_{i_\alpha}^\dagger(\Omega)$ by $c_{i_\alpha}^\dagger$, $c_{i_\alpha}(-\Omega)$ by $c_{i_\alpha}$, leads to the following local Lindblad equation for the OPDM:
\begin{equation}
\frac{d\rho^{(1)}}{dt}=-i[h+h_{LS},\rho^{(1)}]+\sum_\alpha J^\alpha\{(\rho_\alpha^{(1)}-\rho^{(1)}), P_{i_\alpha}\} \label{LindLocal}
\end{equation}
where the Lamb shift Hamiltonian is $h_{LS}=-\sum_\alpha \text{P}\int_{\infty}^{\infty} d\omega \frac{N^\alpha(\omega)}{\omega-\mu_{i_\alpha}} P_{i_\alpha}$ and $\rho_\alpha^{(1)}=f^{\alpha}(\mu_{i_\alpha})I$. Another way to arrive at this result is to notice that in the simple model bath, the bath relaxation time is of order $1/k_B T$ if the reservoir is in the wide-band limit \cite{purkayastha2016}, and the temperature scale $k_B T$ is much larger than the bandwidth of the chain $2t_S$.  In this limit all of the system energy levels may be considered to be the same and Eq. (\ref{LindLocal}) follows directly from Eq. (\ref{ODDMRedfield1}).

\section{Non-equilibrium Green's function approach}\label{SecNEGF}

We review the Keldysh NEGF formalism used later. For concreteness again consider the one-dimensional chain described in Section \ref{FullRedSec}. The main expressions needed to describe transport are described. Similar expressions for the three-dimensional case can be found in Appendix \ref{AppGreen3D}. 

In the stationary state that we are interested in, time-dependent Greens functions $G(t,t^\prime)$ are a function only of the time difference $t - t^\prime$, so the Fourier transform can be expressed in terms of a single frequency: $G=G(\omega)$. The Green's functions of the chain coupled to the baths needed to evaluate observables can be obtained from Dyson's equation:
\begin{eqnarray}
G^{r,a}(\omega)&=&(\omega \pm i \eta -H_S^0 -\Sigma^{r,a}(\omega))^{-1},
\nonumber \\
G^<(\omega)&=&G^r(\omega) \Sigma^<(\omega) G^a(\omega),
\label{eq:keldysh}
\end{eqnarray}
where $G^{r,a,<}(\omega)$ and $\Sigma^{r,a,<}(\omega)$ denote the retarded, advanced, and lesser Green's functions and self-energies respectively that describe  
the coupling of the chain to the baths. The self-energies read:
\begin{eqnarray}
\Sigma_{ij}^{r,a}(\omega) &=& \mp i (J^L  \delta_{i1} \delta_{j1} +J^R  \delta_{iN} \delta_{jN})
\nonumber \\
\Sigma_{ij}^<(\omega)&=&2i\left( J^L f^L(\omega) \delta_{i1} \delta_{j1}   + J^R f^R(\omega) \delta_{iN} \delta_{jN} \right),
\label{eq:selfener}
\end{eqnarray}
where we again take the wide band approximation for the leads so that $J^\alpha=\pi N^\alpha(\epsilon_\lambda)$ is energy independent. The particle current in term of the Green's function is well known and given by: \cite{Yigal1992}
\begin{equation}
j=4 J^L J^R \int_{-\infty}^\infty \frac{d \omega}{2 \pi} \left(f^L(\omega)-f^R(\omega)\right) |G^r_{1N}(\omega)|^2,\label{eq:current1}
\end{equation}  
and the occupation along the chain reads:
\begin{equation}
\langle n_{i} \rangle =- iG^<_{ii}(0)=-i \int_{-\infty}^{\infty} \frac{d \omega}{2 \pi} G^<_{ii}(\omega).
\label{eq:ni}
\end{equation}
The occupation along the chain can be obtained from Eqs. (\ref{eq:keldysh})-(\ref{eq:selfener}) and Eq. (\ref{eq:ni}):
\begin{multline}
\langle n_{i} \rangle =2\int_{-\infty}^{\infty} \frac{d \omega}{2 \pi} \left( J^L f^L(\omega) |G^r_{i1}(\omega)|^2\right. \\ \left.+  J^R f^R(\omega) |G^r_{iN}(\omega)|^2\right).
\label{eq:nif}
\end{multline}
Within the Hartree approximation we could in principle obtain all of the Green's functions by solving Eqs.(\ref{eq:keldysh}) and (\ref{eq:nif}) iteratively, but a more accurate and efficient numerical evaluation of Eq. (\ref{eq:nif}) can be achieved \cite{Ordejon2002} by splitting the integration into the equilibrium and non-equilibrium contributions:
 \begin{eqnarray}
 \langle n_{i} \rangle= \langle n^{L,eq}_{i}  \rangle +  \langle n^{R,neq}_{i} \rangle,
 \nonumber \\
  \langle n_{i} \rangle= \langle n^{R,eq}_{i} \rangle + \langle n^{L,neq}_{i} \rangle,
 \end{eqnarray}
 where the equilibrium part reads:
 \begin{equation}
 \langle n^{\alpha,eq}_{i} \rangle =- \int_{-\infty}^{\infty} {d \omega \over \pi} Im G^r_{ii}(\omega) f^\alpha(\omega),
 \end{equation}
 while the non-equilibrium contributions are:
 \begin{widetext}
  \begin{eqnarray}
 \langle n^{R,neq}_{i} \rangle = J^R \int_{-\infty}^{\infty} {d \omega \over 2 \pi} (f^R(\omega) -f^L(\omega)) |G^r_{iN}(\omega)|^2,
 \nonumber \\
 \langle n^{L,neq}_{i} \rangle = J^L \int_{-\infty}^{\infty} {d \omega \over 2 \pi} (f^L(\omega) -f^R(\omega)) |G^r_{i1}(\omega)|^2.
 \nonumber \\
 \end{eqnarray}
 \end{widetext}

Here, we evaluate the equilibrium contribution using Matsubara sums. From the definition of the
Matsubara Greens function we find:
\begin{eqnarray}
\langle n^{\alpha,eq}_{i,\sigma}\rangle&=&G^{\alpha}_{ii}(0^-)={1 \over \beta} \sum_n e^{i \omega_n 0^+} G^{\alpha}_{ii}(i\omega_n)=
\nonumber \\
&=& {2 \over \beta } \sum_{\omega_n \geq 0} Re G^{\alpha}_{ii}(i\omega_n)+ {1 \over 2}.
\end{eqnarray}
where $G^{\alpha}_{i,i,\sigma}(i\omega_n)$ is the Matsubara Green's function evaluated with the chemical potential of either lead, $\mu=\mu^\alpha$. The 
$\omega_n=0$ contribution in the last Matsubara sum is weighted by a factor $1/2$. The integrals appearing in the non-equilibrium part of the occupation as 
well as in the current Eq. (\ref{eq:current1}) are performed using Gaussian quadrature. 
 
It is enlightening to illustrate the connection between MRE and NEGF here. QME can also be formally derived using Keldysh diagrammatic technique, by carefully picking certain relevant diagrams \cite{Kleinherbers2020}. Here we make the connection from the perspective of Kadanoff-Baym equation \cite{Kadanoff1962}.  The equation of motion version of the Keldysh formalism, as introduced by Kadanoff and Baym, leads to an equation for $G^<(t,t^\prime)$ which reads \cite{Jauho}:
\begin{eqnarray}
&&i(\partial_t +\partial_{t^\prime})G^<(t,t^\prime)-[H_S^0,G^<(t,t^\prime)] =
\nonumber \\
&=& \int d t_1 (\Sigma^r(t,t_1) G^<(t_1,t^\prime)  + \Sigma^<(t,t_1) G^a(t_1,t^\prime) 
\nonumber \\
&-& G^r(t,t_1) \Sigma^<(t_1,t^\prime) - G^<(t,t_1) \Sigma^a(t_1,t^\prime)),
\label{eq:eom}
\end{eqnarray} 
in matrix form. Note that in order to solve for $G^<(t,t^\prime)$, the time dependence of $G^{r,a}(t,t^\prime)$ is also needed. This dependence  
can be generally obtained by solving the Dyson equation of motion in integro-differential form for $G^{r,a}(t,t^\prime)$. We now apply the above expression (\ref{eq:eom}) to the same non-interacting one-dimensional tight-binding model used before. 
Equation (\ref{eq:eom}) above simplifies in the stationary regime in which not only the
baths (which are in equilibrium) but also all the Keldysh Green's functions 
depend only on the time-difference $G(t,t^\prime)=G(t-t^\prime)$. After Fourier transforming the Greens functions and setting $t=t^\prime$ we obtain:
\begin{eqnarray}
&&-i\sum_k (H_S^0)_{ik} \rho_{kj}^{(1)} + i\sum_k \rho_{ik}^{(1)} (H_S^0)_{kj} -J_L \delta_{i1} \rho_{1j}^{(1)} 
\nonumber \\
&&-J_R \delta_{iN} \rho_{Nj}^{(1)} - \rho_{i1}^{(1)} J_L \delta_{j1} - \rho_{iN}^{(1)} J_R \delta_{jN}
\nonumber \\
&=&2i\int {d \omega \over 2 \pi} \left(  J_L f_L(\omega) \delta_{i1}  G^a_{1j} (\omega) +  J_R f_R(\omega) \delta_{iN}  G^a_{Nj}(\omega) \right)
\nonumber \\
&-&2i\int {d \omega \over 2 \pi} \left( G^r_{i1}(\omega)  J_L f_L(\omega) \delta_{j1}+  G^r_{iN} (\omega)  J_R f_R(\omega) \delta_{jN}  \right),
\nonumber \\
\label{negfsteady}
\end{eqnarray} 
where we have used the definition: $\langle c^\dagger_i c_j \rangle = -i G^<_{ji}(t,t)=\rho_{ji}^{(1)}$. Note that
for non-interacting electrons in the absence of baths, the retarded and advanced Greens function can be expressed in terms 
of the eigenvalues and eigenvectors $\{ \epsilon_\lambda, | \lambda \rangle \}$ of the system Hamiltonian $H_S^0$ as:
\begin{equation}
G_{ij}^{r,a}(\omega)=P\sum_{\lambda} {\langle i | \lambda \rangle \langle \lambda | j \rangle \over \omega - \epsilon_\lambda} \mp i \pi \sum_\lambda \langle i | \lambda \rangle \langle \lambda | j \rangle \delta(\omega -\epsilon_\lambda). 
\end{equation}
Substituting this expression into Eq. (\ref{negfsteady}) leads to an equation equivalent to Eq. (\ref{ODDMRedfield1}) that was obtained within the MRE approach without neglecting the CPVPs (note that we have implicitly assumed $\hat{K}^\alpha=0$ in Eq. (\ref{eq:selfener})). Therefore, our MRE approach effectively ignores the direct contribution from the system-bath coupling to $G^{r,a}(\omega)$ in Eq. (\ref{negfsteady}). A similar argument shows that the implementation of the mean-field approximation within MRE is equivalent to replacing all the explicit self-energies in Eq. (\ref{eq:eom}) with the mean-field self-energies plus the system-bath self-energies given by Eq. (\ref{eq:selfener}), and only keeping the self-consistent mean-field self-energies in Eq. (\ref{eq:keldysh}). 
Note that, due to the wide band approximation, the instantaneous self-energy $\Sigma^r(t,t^\prime) \propto \delta(t-t^\prime)$, which 
when substituted into Eq. (\ref{eq:eom}) leads to local-in-time dynamics. Non-Markovian dynamics may, however, arise as a result of the edge
singularities present in generic fermionic baths. \cite{chakraborty2018}

\section{Application to specific systems}\label{SecApp}

\subsection{One-dimensional ballistic transport in non-interacting metallic quantum wire}

We first study the simplest case of a spinless non-interacting chain of fermions that is connected to reservoirs at each end as shown in Fig. \ref{Fig1}. The system Hamiltonian is given by Eq. (\ref{sysH}) and for simplicity 
we set $\mu_i = V_{i,\sigma;j,\sigma^\prime} \equiv 0$, and $J^\alpha \equiv J$. Eq. (\ref{ODDMRedfield2}) shows that, as the system approaches the steady state, in real space the OPDM obeys
\begin{multline}
it_S(\rho_{2n+1,1}^{(1)}+\rho_{n,n}^{(1)}-\rho_{n+1,n+1}^{(1)})=J(\rho^{(1)}-\rho_{L}^{(1)})_{2n,1}, \label{Ball1}
\end{multline}
with particle current $j$ equal to
\begin{equation}
j=it_S(\rho_{2,1}^{(1)}-\rho_{1,2}^{(1)})=2J(\rho^{(1)}-\rho_{L}^{(1)})_{1,1} \label{jL1}
\end{equation}
where $1\leq n\leq (N-1)/2$. For the uniform chain, we expect that the interior density $\rho_{n,n}^{(1)}$ asymptotically approaches a fixed value $\rho_b$. Likewise $\rho_{2n+1,1}^{(1)}=\langle c_1^\dagger c_{2n+1}\rangle$ should be small when $n \gg 1$. Summing over $n$ on both sides of Eq. (\ref{Ball1}), keeping the imaginary parts, and dropping  $\rho_{m,1}^{(1)}$ when $m\geq 3$, yields
\begin{equation}
j=2t_S\text{Im}\ \rho_{2,1}^{(1)}=\frac{2t_S^2}{J}(\rho_{1,1}^{(1)}-\rho_b). \label{jL2}
\end{equation}
Similarly we have
\begin{equation}
j=it_S(\rho_{N-1,N}^{(1)}-\rho_{N,N-1}^{(1)})=2J(\rho^{(1)}-\rho_{R}^{(1)})_{N,N} \label{jR1}
\end{equation}
and
\begin{equation}
j=2t_S\text{Im}\ \rho_{N,N-1}^{(1)}=-\frac{2t_S^2}{J}(\rho_{N,N}^{(1)}-\rho_b) \label{jR2}
\end{equation}
where we use the fact that a system at equilibrium has no current. Combining Eqs.(\ref{jL1}), (\ref{jL2}), (\ref{jR1}), and (\ref{jR2}) we obtain
\begin{equation}
\rho_b=\frac{(\rho^{(1)}_{L})_{1,1}+(\rho^{(1)}_R)_{N,N}}{2}, \label{bulkrho}
\end{equation}
and
\begin{equation}
j=\frac{J}{1+J^2/t_S^2}\left((\rho^{(1)}_L)_{1,1}-(\rho^{(1)}_R)_{N,N}\right)\ . \label{Current1}
\end{equation}
Alternatively we may write Eq. (\ref{Current1}) as
\begin{multline}
j=\frac{J}{1+J^2/t_S^2}\sum_E \Gamma(E)\left( f^L(E)-f^R(E) \right) \\=\frac{J}{1+J^2/t_S^2}\int_{-\infty}^{\infty} dE N(E)\Gamma(E)\left( f^L(E)-f^R(E) \right)\\
\approx\frac{J/t_S}{\pi(1+J^2/t_S^2)}\int_{-\infty}^{\infty} dE \sqrt{1-E^2/4t_S^2} \left( f^L(E)-f^R(E) \right) \label{Current2}
\end{multline}
where $\Gamma(E)=\vert \langle 1\vert E \rangle\vert^2=2(1-E^2/4t_S^2)/(N+1)$, $N(E)=N/(2\pi t_S\sqrt{1-E^2/4t_S^2})$ is the density of states of the system and we have assumed thermodynamic limit. 
\begin{figure}[t]
\centering
  \subfigure[]{
    \label{1d:1} 
    \includegraphics[width=2.6 in]{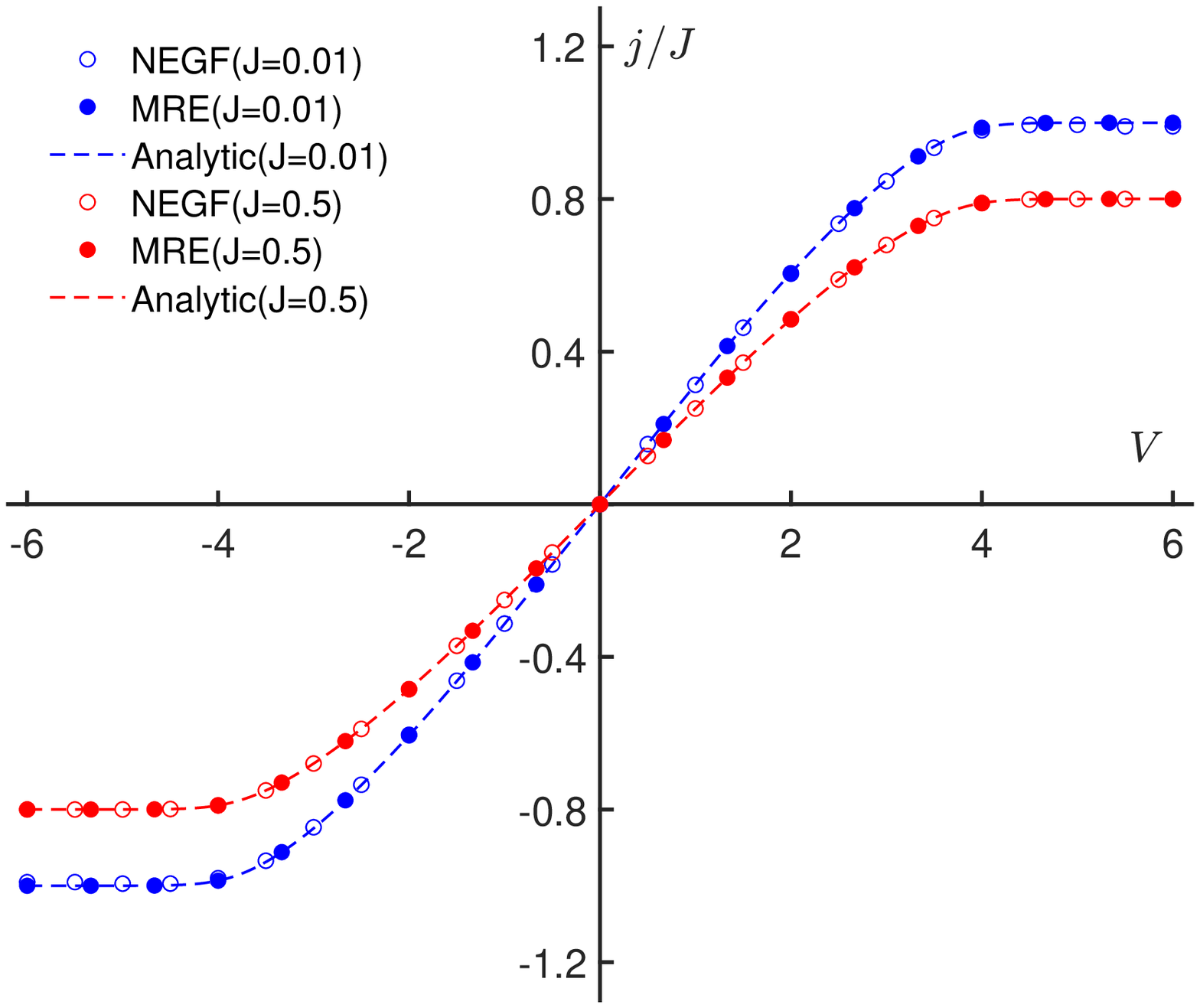}}

  \subfigure[]{
    \label{1d:2} 
    \includegraphics[width=2.6 in]{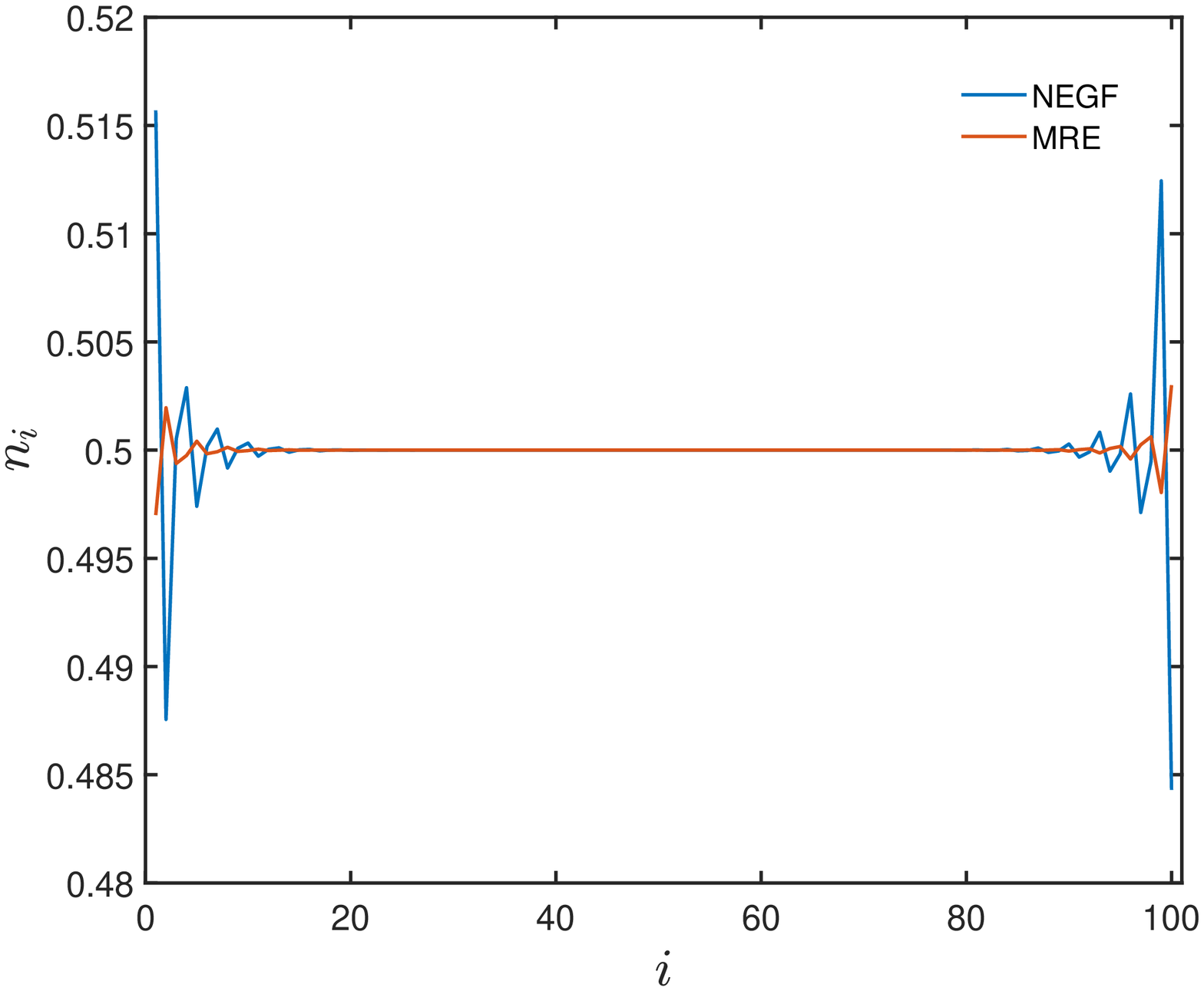}}
\caption{(a) Current-voltage I-V curve of the one dimensional conducting quantum wire for  different $J$, calculated by NEGF, MRE and Eq. (\ref{Current2}). (b) Occupancy profile when $J=0.1$ and $V=2$, calculated by NEGF and MRE methods. Voltage is defined as $V=\mu^R-\mu^L$. For all plots the parameters chosen are: $t=1$, $T=0.1$, $N=100$, $\mu^R+\mu^L=0$. }
\label{1d}
\end{figure}
As the Redfield equations are based on the weak-coupling assumption \cite{fleming2011}, care must be taken in using Eq. (\ref{Current2}) when $J$ is large. Eq. (\ref{bulkrho}) can be interpreted with  Landauer ballistic transport as Eq. (\ref{Current2}) is similar to Landauer's formula $j_L=\int_{-\infty}^{\infty} dE\left( f^L(E)-f^R(E) \right)/2\pi$ for only one conducting channel \cite{landauer1970}; however, different energy levels contribute differently to the current and the total current is suppressed by a factor of $\sim J/(1+J^2/t_S^2)$. This difference originates from the fact that the system-bath interface in our model can reflect electrons that travel from the system to the bath. Fig. \ref{1d:1} shows excellent agreement between numerical results obtained by RME, NEGF, and the analytic formula of Eq. (\ref{Current2}). As shown in Fig. \ref{1d:2}, the occupancy determined by MRE and NEGF differs at the boundary of the system but in the bulk of the system the difference between two methods is negligible. These results suggest that RME is reliable for the investigation of transport properties in the thermodynamic limit.

\subsection{Semiconductor p-n junctions under DC bias}\label{SecPn}
\begin{figure*}[!ht]
\centering
  \subfigure[]{
    \label{fccmodel:1} 
    \includegraphics[width=2.5 in]{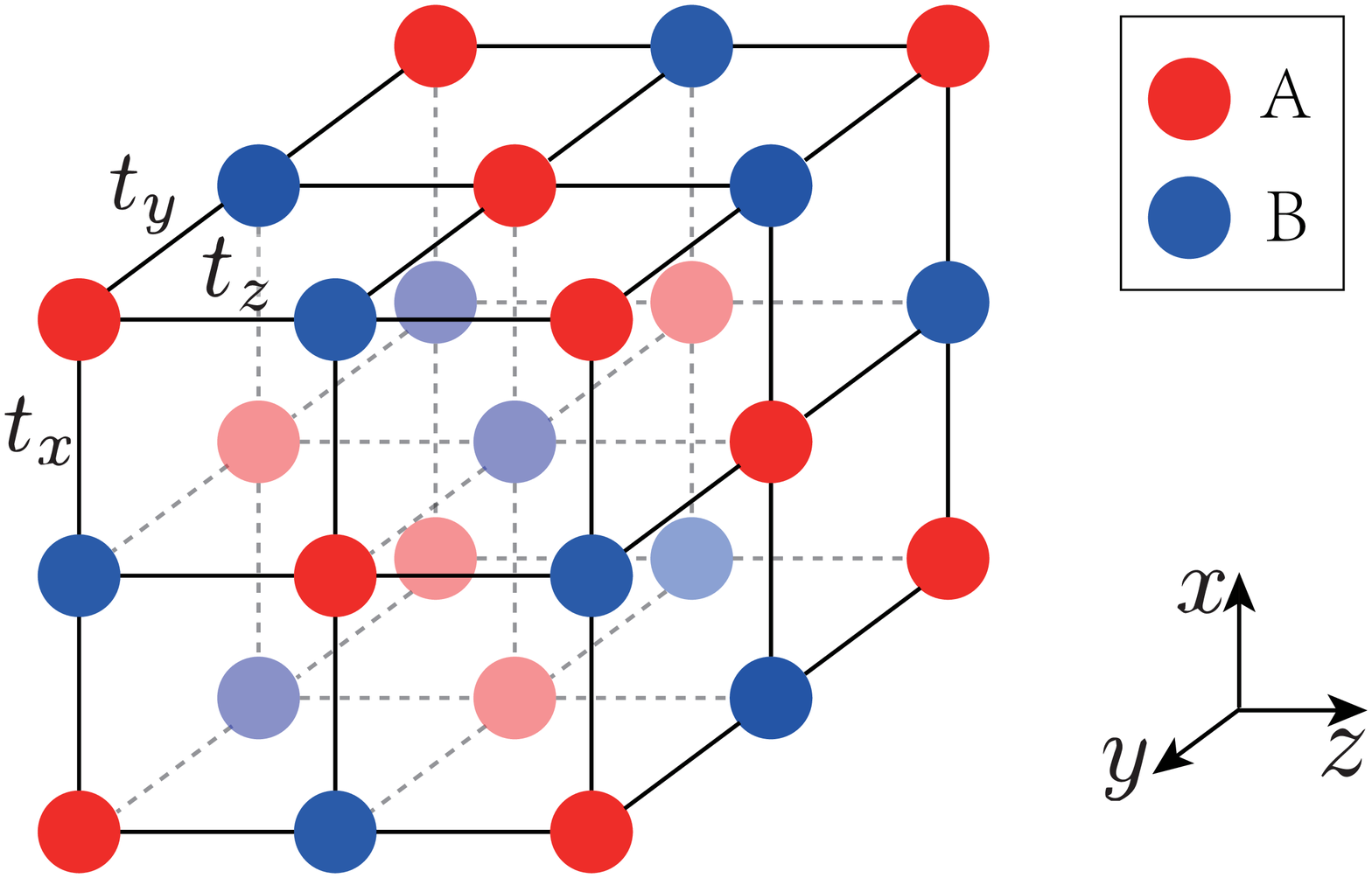}}
  \hspace{0.5in}
  \subfigure[]{
    \label{fccmodel:2} 
    \includegraphics[width=1.8 in]{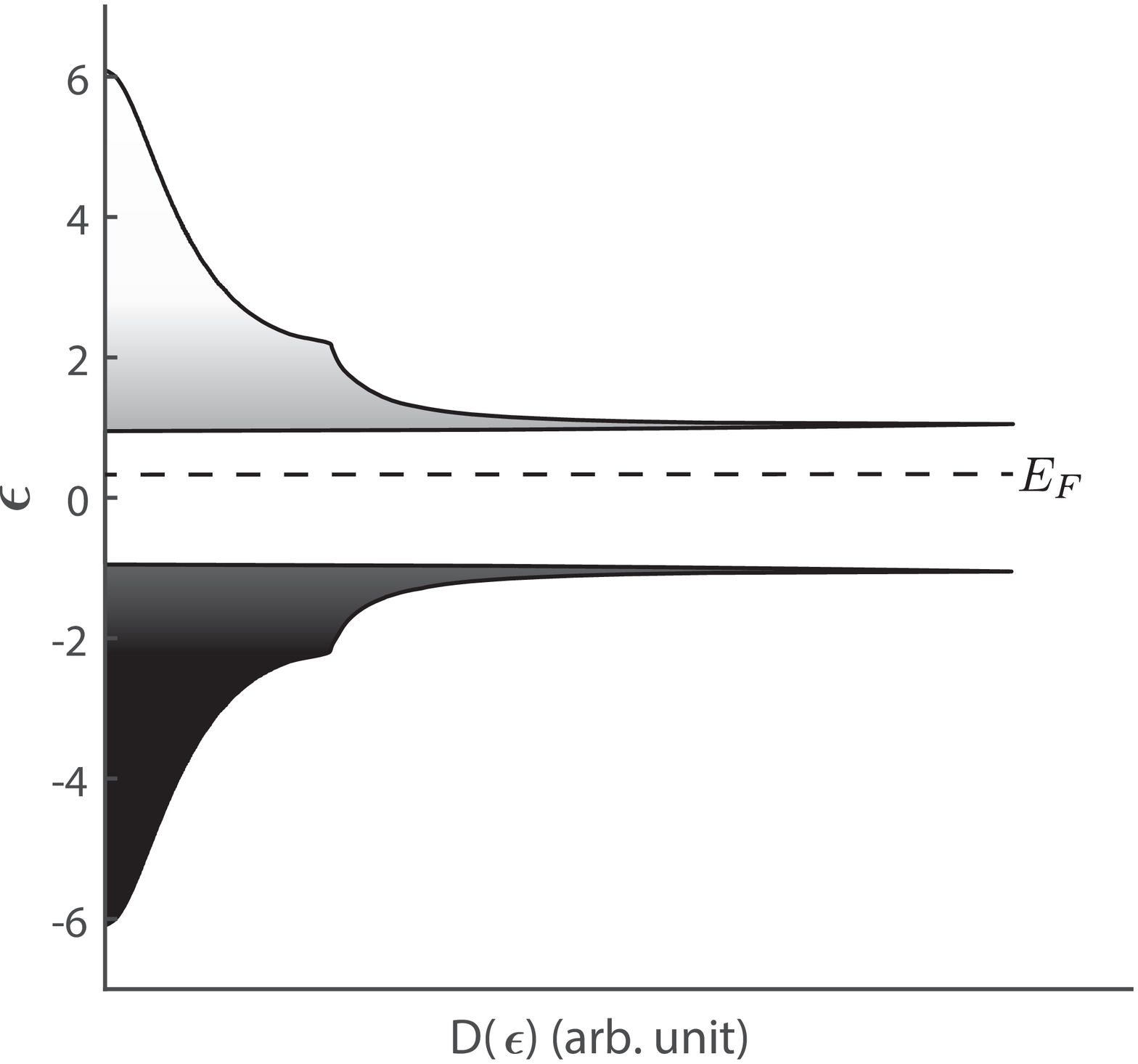}}
\caption{(a) Lattice structure of the three-dimensional semiconductor. (b) Density of states $D(\epsilon)$ in this model.}
\label{fccmodel}
\end{figure*}
\begin{figure}[!ht]
\includegraphics[width=3 in]{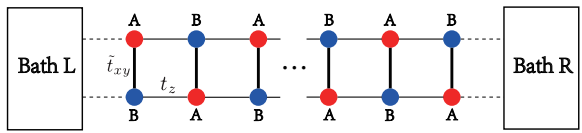}
\caption{Effective one-dimensional tight-binding model with ladder geometry.  The hopping matrix element $\tilde{t}_{xy}$ depends on $k_x$ and $k_y$.}
\label{Effmodel}
\end{figure}
Next we consider the three-dimensional face-center-cubic (FCC) tight-binding model illustrated in Fig. \ref{fccmodel}.
Hopping is turned on only between nearest-neighbor sites and thus connects sublattice A to B.
To open a gap, alternating on-site energies of $\pm \Delta$ are assigned to the two sublattices.  Because the system has translational invariance along the two transverse directions, for each different wavevector $\mathbf{k}=(k_x, k_y)$ the three-dimensional problem is described by an effective 1-D tight-binding ladder model, as sketched in Fig. \ref{Effmodel}.
\begin{figure*}[!ht]
\centering
  \subfigure[]{
    \label{3dpn:1} 
    \includegraphics[width=2.6 in]{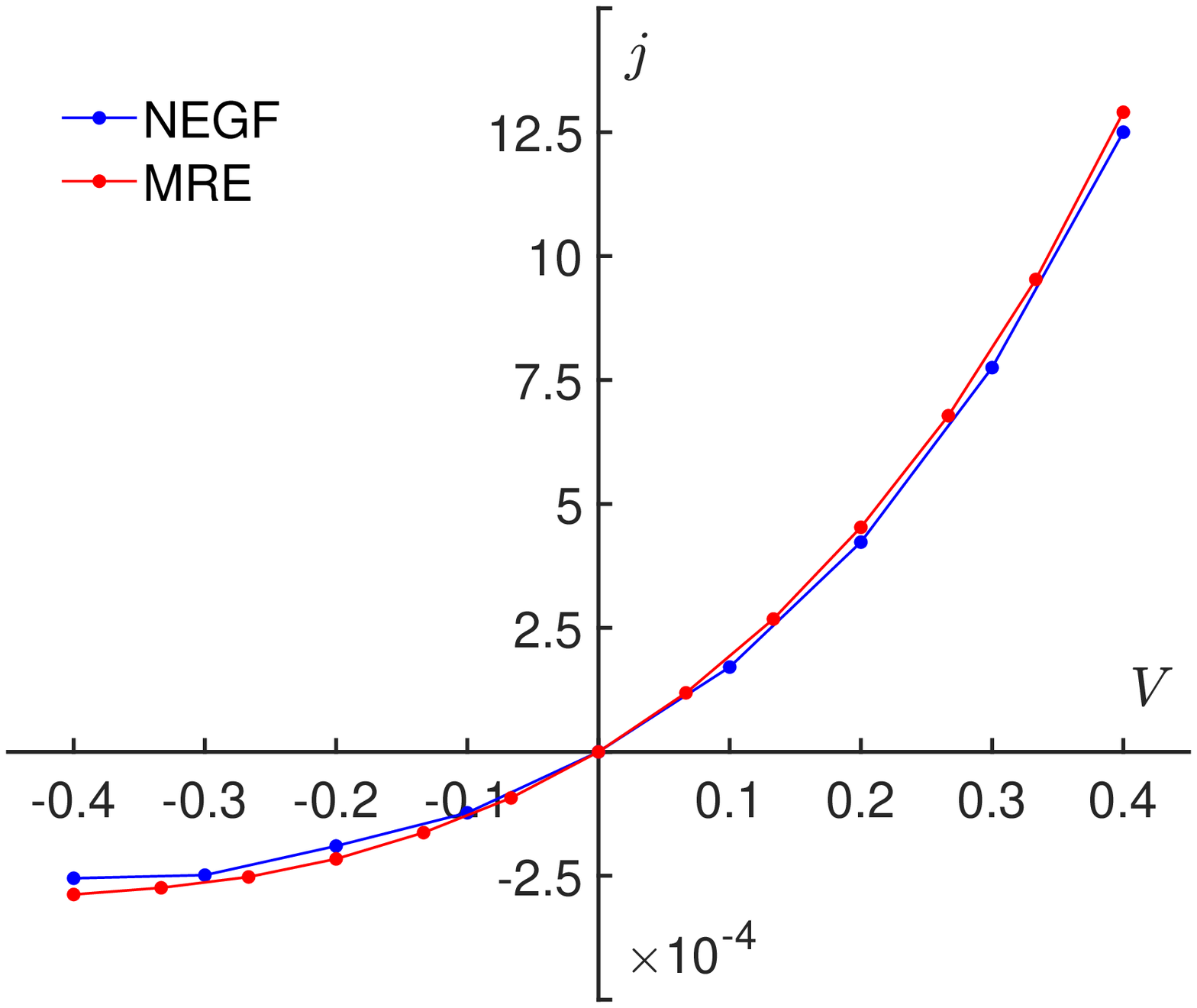}}

  \subfigure[]{
    \label{3dpn:2} 
    \includegraphics[width=2.6 in]{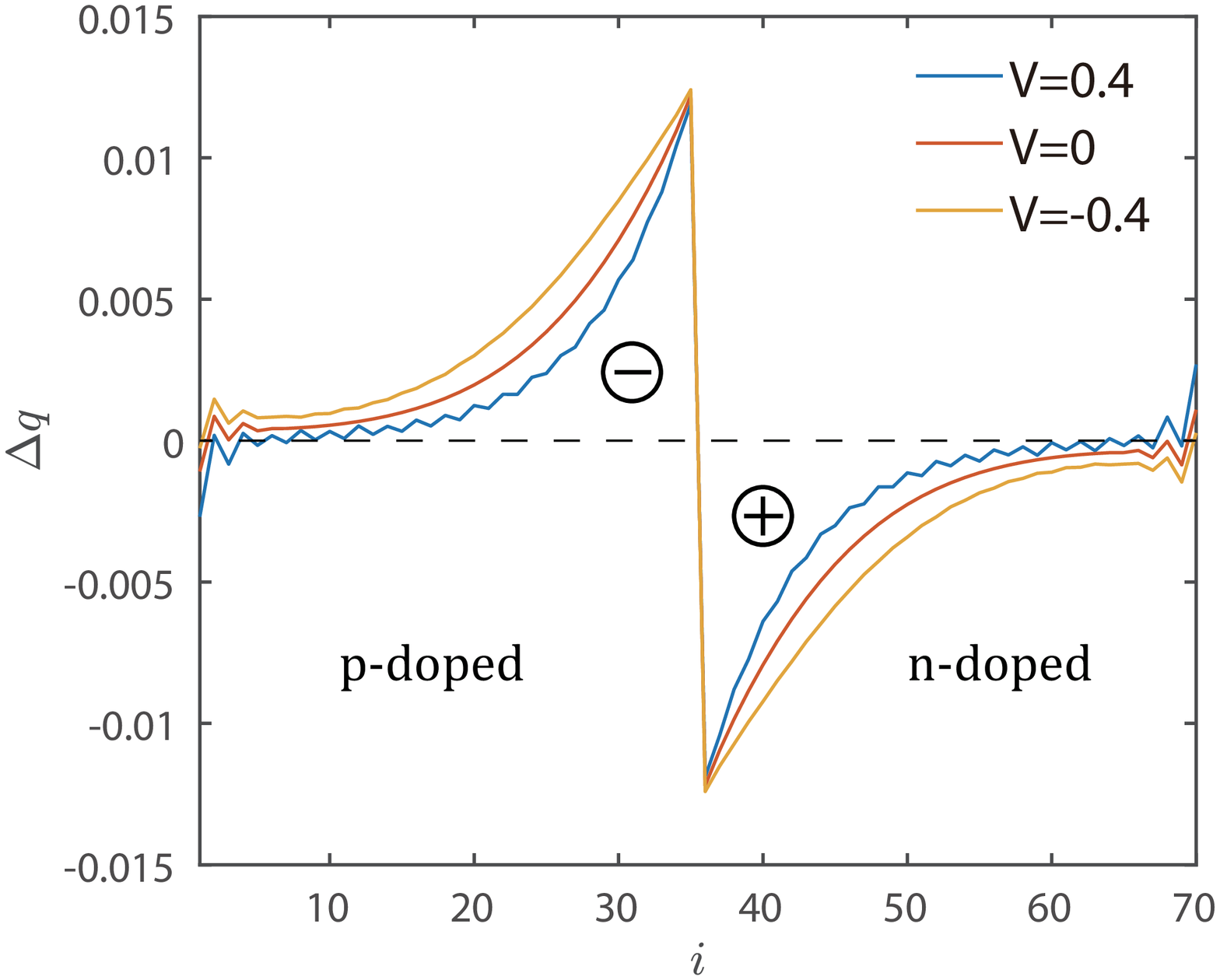}}
  \hspace{0in}
  \subfigure[]{
    \label{3dpn:3} 
    \includegraphics[width=2.6 in]{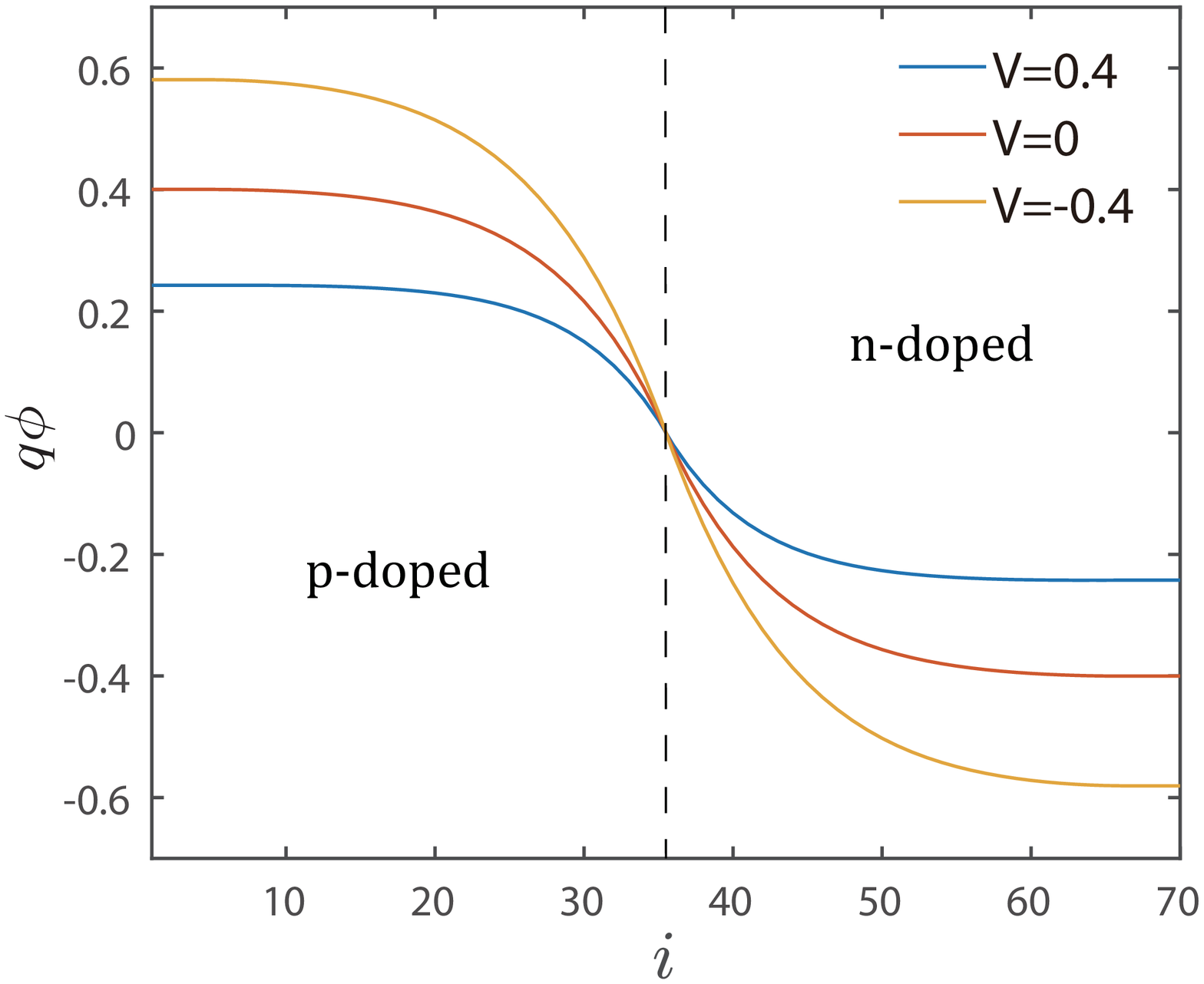}}
\caption{(a) Current-voltage I-V curve of the three dimensional semiconductor p-n junction from the MRE and NEGF methods. (b) Excess negative charge per site along the $z$ direction under different biases as  calculated with MRE. (c)The electrostatic potential energy profile along the $z$ direction under different bias calculated with MRE. The bias voltage is defined as $V=\mu^R-\mu^L$ and $\mu^R+\mu^L=0$ is chosen to maintain charge neutrality. For all the plots the parameters are chosen as: $N_x=N_y=8$, $N_z=70$ (number of unit cells along $z$ direction), $T=0.3$, $V_0=0.02$, $J=0.5$, $\overline{n}_{L(R)}=0.974\ (1.026)$  (number of background charge on left (right) side), $t_z=t_x=t_y=\Delta=1$. Note that for the MRE results, oscillating charge within the 4 unit cells closest to the boundaries is excluded from the calculation of the electrostatic potential.  We have checked that this does not affect the results appreciably.}
\label{3dpn}
\end{figure*}
The one-dimensional Hamiltonian is
\begin{multline}
h_{\mathbf{k}}=-\sum_i \left[ t_z\left( c_{i,A}^\dagger c_{i+1,B}+c_{i,B}^\dagger c_{i+1,A}\right) \right. \\
+\left.  \tilde{t}_{xy} c_{i,A}^\dagger c_{i,B}+\text{H.c.}\right]+\sum_i\Delta (n_{i,A}-n_{i,B})  \label{1DTB}
\end{multline} 
with $\tilde{t}_{xy}=2t_x\cos(k_x)+2t_y\cos(k_y)$ and we have indicated index $\mathbf{k}$ implicitly. To model p-n junctions the positive charge background must be included as well as the long-range Coulomb interaction as these ingredients are necessary to model the depletion layer of a standard semiconductor p-n junction. Here we make the Hartree approximation with a mean-field Hamiltonian that ignores correlation and exchange:
\begin{equation}
H_{int}^{MF}=\sum_{\mathbf{R}\neq\mathbf{R^\prime}}\sum_\sigma 2V_{\mathbf{R}\mathbf{R^\prime}}\left( \langle n_{\mathbf{R^\prime},\sigma}\rangle-\frac{1}{2}\overline{n}_\mathbf{R^\prime}\right)  n_{\mathbf{R},\sigma}
\end{equation}
where $\overline{n}_\mathbf{R}$ denotes the charge of the positive charge background at site $\mathbf{R}$, and $V_{\mathbf{R}\mathbf{R^\prime}}=\frac{V_0}{\vert\mathbf{R}-\mathbf{R^\prime}\vert}$ is the Coulomb interaction. We have restored the index of spin $\sigma$ for clarity, and take advantage of spin rotational symmetry. For an intrinsic semiconductor without doping, the system is at half-filling, so we set $\overline{n}_\mathbf{R^\prime}=1$ for all $\mathbf{R^\prime}$. By tuning $\overline{n}_\mathbf{R^\prime}$ to be slightly above or below unity we can model either a n-doped or a p-doped semiconductor. Due to the translational symmetry along $x$ and $y$ directions, it is possible to further simplify the Hamiltonian to 
\begin{equation}
H_{int}^{MF}=\sum_{i,j,\mathbf{R_{\perp}}}\sum_{\sigma}\tilde{V}_{ij}\left( \langle n_{j,\sigma}\rangle-\frac{1}{2}\overline{n}_j\right) n_{i,\mathbf{R_{\perp}},\sigma} \label{1Dinteraction}
\end{equation}
where $i,j$ are the indices labelling the coordinate of site along $z$ direction, $\mathbf{R_{\perp}}$ labels the coordinate along $x$ and $y$ direction, $\langle n_{j,\sigma}\rangle=\langle n_{j,A,\sigma}+n_{j,B,\sigma}\rangle/2$ and $\tilde{V}_{ij}\approx-4\pi V_0\vert i-j\vert$, which is obtained by approximately replacing $\sum_{\mathbf{R^\prime}}V_{\mathbf{R}\mathbf{R^\prime}}$ with an integral and ignoring the overall energy shift term. Transforming Eq. (\ref{1Dinteraction}) to $k$-space we obtain the effective 1-D interaction Hamiltonian for each different $\mathbf{k}$:
\begin{equation}
h^{int}_{\mathbf{k}}=\sum_{i,j,s,\sigma}\tilde{V}_{ij}\left( \sum_{\mathbf{k^\prime}}\frac{\langle n_{\mathbf{k^\prime},j,\sigma}\rangle}{N_xN_y}-\frac{1}{2}\overline{n}_j\right)n_{\mathbf{k},i,s,\sigma} \label{final1Dint}
\end{equation} 
where $N_x$ and $N_y$ are the number of unit cells along $x$ and $y$ direction respectively, $s$ is the index labelling sublattice $A$ and $B$, and $\langle n_{\mathbf{k},j,\sigma}\rangle=\langle n_{\mathbf{k},j,A,\sigma}+n_{\mathbf{k},j,B,\sigma}\rangle/2$. By setting $\overline{n}_j$ differently on the left and right halves of the ladder, a semiconductor p-n junction can be modeled. 

The corresponding MRE of the OPDM for the 3D system is:
\begin{multline}
\frac{d\rho^{(1)}_{\mathbf{k}}}{dt}=-i[h_{\mathbf{k}}+h^{int}_{\mathbf{k}},\rho_{\mathbf{k}}^{(1)}]\\+J\sum_{\alpha}\{\rho_{\alpha,\mathbf{k}}^{(1)}-\rho_{\mathbf{k}}^{(1)},\sum_{s}P_{\mathbf{k},i_\alpha,s}\} \label{1DRedfield1}
\end{multline}
where
\begin{equation}
\rho^{(1)}_{\mathbf{k}}=\sum_{\alpha,\beta}\vert \mathbf{k},\alpha \rangle\text{Tr}(c_{\mathbf{k},\beta}^\dagger c_{\mathbf{k},\alpha}\rho)\langle\mathbf{k},\beta \vert, \label{1DRedfield2}
\end{equation}
\begin{equation}
\rho_{\alpha,\mathbf{k}}^{(1)}=\sum_\lambda \vert \mathbf{k},\lambda\rangle f^\alpha(\epsilon_{\mathbf{k},\lambda})\langle \mathbf{k},\lambda\vert\ . \label{1DRedfield3}
\end{equation}
Here $\vert \mathbf{k},\lambda\rangle$ is the eigenstate of $h_{\mathbf{k}}+h^{int}_{\mathbf{k}}$, and $P_{\mathbf{k},i_\alpha,s}$ projects onto state $\vert \mathbf{k},i_\alpha,s\rangle$. Note that here we have assumed the generalized spectral density defined in Appendix \ref{AppRed3D} to be $N^\alpha(\mathbf{k};i_\alpha,A;i_\alpha,A)=N^\alpha(\mathbf{k};i_\alpha,B;i_\alpha,B)=J$ and vanishes otherwise. Solving Eq. (\ref{1DRedfield1}) self-consistently yields the OPDM of the non-equilibrium steady state.

The I-V curves of the p-n junction obtained from the essentially exact 
NEGF and MRE are shown in Fig. \ref{3dpn:1}.
There is only a small difference between the two approaches, providing support for the MRE approximation. The I-V curves are similar to those of a textbook semiconductor p-n junction and show clear rectification. From Fig. \ref{3dpn:2} and Fig. \ref{3dpn:3}, we see that the size of the depletion region changes when the bias changes as expected. However, since there is no disorder or dissipation mechanism in the bulk of the system, the transport is ballistic everywhere except at the p-n interface and the carriers are in a highly out-of-equilibrium state everywhere.  This contrasts with textbook descriptions of macroscopic p-n junctions where the carriers are in a quasi-equilibrium state away from the p-n interface and a recombination current is important in the vicinity of the junction. We also note from Fig. \ref{3dpn:3} that the shift in the electrostatic potential at the baths does not exactly equal the external bias. This difference is due to the absence of disorder and dissipation in our model.

\section{Conclusion}
\label{Conclusion}
In this paper, we derive the Redfield equation of motion for the one-particle density matrix. By ignoring the Cauchy principle value parts (CPVPs) and making a mean-field approximation, we obtain a closed equation (\ref{ODDMRedfield2}) that can be efficiently solved numerically. Application to the one-dimensional metallic quantum wire system shows that the approach captures the essential ballistic transport physics and leads to the Landauer current formula. One difference with the non-equilbrium Green's function (NEGF) approach is apparent in the occupancy near the boundary with the baths.  The difference has two sources: (1) correlations between the system and bath, which are neglected in general quantum master equations, may be important near the system-bath interface; and (2) the CPVPs that we ignore are localized near the boundaries.  Nonetheless, for mesoscopic and macroscopic systems, transport appears to be only minimally affected (see Fig. \ref{1d:1}).  Application of the modified Redfield equation (MRE) to a three-dimensional semiconductor p-n junctions yields a good match between the MRE and NEGF I-V curves with strong rectification as expected (see Fig. \ref{3dpn}). There are slight differences between our results and those of textbook p-n junctions, as discussed in Section \ref{SecPn}.  MREs serve as an alternative method for the investigation of transport in the inhomogeneous heterostructures \citep{yonemitsu2007,yonemitsu2009}. 

MREs may be better suited to describe transport in mesoscopic systems than the Lindblad equations, as the microscopic derivation of the Lindblad equations requires either a secular approximation, which breaks down, or a high-temperature limit. The range of validity of MREs can be explored by direct comparison to NEGF (see Section \ref{SecNEGF}), showing that the two methods are closely related and in fact almost equivalent in the weak system-bath coupling limit. For mean-field approximations MRE is more numerically more efficient than NEGF as it avoids summations over frequency and the need to solve complicated integro-differential equations.  Positive definiteness of the density matrices is not guaranteed for MREs, however, and positivity needs to be checked on a case-by-case basis. 

The MRE approach introduced here seems to be limited, in practice, to weakly interacting systems that are adequately 
described by a mean-field approach. At the mean-field level,
the set of MREs decouple forming
a closed system of equations that can be solved.
Blind application of MREs to strongly interacting systems would, in principle, require the solution of an infinite hierarchy of equations that couple
correlation functions at different orders. This closure problem may be avoided by truncating and solving the hierarchy of MREs through a coupled cluster expansions up to a desired order in the interaction \cite{Akbari2012}.
Truncated Redfield equations have already been used for analyzing charge transport through 1D chains with interacting spinless fermions and thermal conduction through XXZ spin chains \cite{wu2010} yielding promising results.  It may be possible to study
transport through 3D strongly correlated system systems via the MREs in the limit of infinite dimensions by 
analogy to the NEGF + DMFT approach which neglects non-local spatial correlations.  These theoretical routes are beyond the scope of our present work and deserve further exploration.

\begin{acknowledgments}
Z.Z. thanks K. Ma for early useful discussions. Z.Z. and J.B.M. acknowledge support from NSF grant no. 1936221.  J.M. acknowledges financial support from (RTI2018-098452-B-I00) Ministerio de Ciencia, Innovaci\'on y Universidades/FEDER, Uni\'on Europea.
\end{acknowledgments}

\appendix
\section{Microscopic derivation of the Redfield equation} \label{AppRedfieldDerivation}

Here we provide a heuristic derivation of the Redfield equation. For a more rigorous derivation see, for example, Ref. \cite{rivas2012}. It is usually convenient to treat $H_{SB}=\sum_\alpha H_{SB}^{(\alpha)}$ as an  interaction and work in the interaction picture. Then the von Neumann equation becomes
\begin{equation}
\frac{d\rho^I_{\text{total}}(t)}{dt}=-i[H_{SB}^I(t),\rho^I_{\text{total}}(t)] \label{VonNeumannInt}
\end{equation}
where 
\begin{equation}
H_{SB}^I(t)=U^\dagger (t,0)H_{SB}^SU(t,0),
\end{equation}
\begin{equation}
\rho^I_{\text{total}}(t)=U^\dagger (t,0)\rho^S_{\text{total}}(t)U(t,0)
\end{equation}
where $U(t_1,t_0)=e^{-i\mathcal{T}\int_{t_0}^{t_1}(H_S(t^\prime)+H_B(t^\prime))dt^\prime}$ is the evolution operator, $H^S_{SB}$ is the system-bath coupling Hamiltonian in the Schr\"odinger picture, $\mathcal{T}$ is the time-ordering operator, $\rho^S_{\text{total}}$ is the total density matrix in the Schr\"odinger picture and initially $\rho_{\text{total}}(0)=\rho_S(0)\otimes \rho_E(0)$. With Eq. (\ref{VonNeumannInt}) one can obtain
\begin{multline}
\rho^I_{\text{total}}(t)=\rho^I_{\text{total}}(0)+\frac{1}{i}\int_0^t dt_1[H_{SB}^I(t_1),\rho^I_{\text{total}}(0)]\\+\frac{1}{i^2}\int_0^t dt_1 \int_0^{t_1}  dt_2 [H_{SB}^I(t_1),[H_{SB}^I(t_2),\rho^I_{\text{total}}(t_2)]] \label{2ndexpansion}
\end{multline}
Assuming that the coupling between bath and system is weak and the bath evolves only slowly (the Born approximation) the density matrix can be written as $\rho_{\text{total}}^I(t)\approx\rho^I_S(t)\otimes \rho^I_E(0)$, therefore after tracing out the degree of freedom of bath and taking derivative with respect to $t$, Eq. (\ref{2ndexpansion}) becomes
\begin{multline}
\frac{d\rho^I_S}{dt}=-\sum_{\alpha}\int_{0}^t dt^\prime\left( [c_{i_\alpha}(t),\hat{C}^{\alpha+}(t,t^\prime)\rho^I_S(t^\prime)]\right. \\+\left. [c_{i_\alpha}^\dagger(t),\hat{C}^{\alpha-}(t,t^\prime)\rho^I_S(t^\prime)]+\text{H.c.}\right). \label{drhodt}
\end{multline}
where
\begin{equation}
\hat{C}^{\alpha+}(t,t^\prime)=\sum_\lambda \vert T_\lambda^\alpha\vert^2\langle f_\lambda^{\alpha\dagger}(t)f^\alpha_\lambda(t^\prime)\rangle_B c_{i_\alpha}^\dagger(t^\prime), \label{C+}
\end{equation}
\begin{equation}
\hat{C}^{\alpha-}(t,t^\prime)=\sum_\lambda \vert T_\lambda^\alpha\vert^2\langle f^\alpha_\lambda(t)f_\lambda^{\alpha\dagger}(t^\prime)\rangle_B c_{i_\alpha}(t^\prime) \label{C-}
\end{equation}
where we have used the fact that $\langle f^\alpha_\lambda\rangle_B=\langle f^{\alpha\dagger}_\lambda\rangle_B=0$, and $\langle ...\rangle_B$ denotes the average value taken with respect to the bath. Here we will assume that the system time scale $\tau_S$ is much larger than the baths' relaxation time scale $\tau_B$ so that at time scale $\tau_B\ll\tau\ll \tau_S$ the dynamics of the coarse-grained density matrix of the system becomes local in time, which is known as the Markov approximation. The evolution of the system does not depend its history, and the bath is memoryless. Therefore, we can replace all the $\rho^I_S(t^\prime)$ by $\rho^I_S(t)$ in Eq. (\ref{drhodt}) and extend the lower limit in the integrand of Eq. (\ref{drhodt}) to $-\infty$. Introducing the concept of eigenoperator defined in Eq. (\ref{eigOp}), one can easily show that
\begin{equation}
e^{iH^S t^\prime} [c_{i_\alpha}(\Omega)]^\dagger e^{-iH^S t^\prime}=e^{i\Omega t^\prime}[c_{i_\alpha}(\Omega)]^\dagger, \label{EOMCdagger}
\end{equation}
\begin{equation}
e^{iH^S t^\prime} c_{i_\alpha}(\Omega)e^{-iH^S t^\prime}=e^{-i\Omega t^\prime}c_{i_\alpha}(\Omega). \label{EOMc}
\end{equation}

Using the fact that
\begin{equation}
\langle f_\lambda^{\alpha\dagger}(t) f_{\lambda}^\alpha(t^\prime)\rangle_B=e^{i\epsilon^\alpha_\lambda(t-t^\prime)}f^\alpha(\epsilon^\alpha_\lambda),
\end{equation} 
\begin{equation}
\langle f_\lambda^\alpha(t) f_{\lambda}^{\alpha\dagger} (t^\prime)\rangle_B=e^{-i\epsilon^\alpha_\lambda(t-t^\prime)}\left(1-f^\alpha(\epsilon^\alpha_\lambda)\right)
\end{equation}
and the definition of spectral density $N^\alpha(\Omega)=\sum_\lambda \vert T_\lambda^\alpha\vert^2\delta(\Omega-\epsilon^\alpha_{\lambda})$, one can obtain
\begin{widetext}
\begin{multline}
\frac{d\rho^I_S}{dt}=-\sum_{\alpha,\Omega,\Omega^\prime}\left\lbrace (F^\alpha(\Omega)+iF^{\alpha I}(\Omega))\times[c_{i_\alpha}(\Omega^\prime),[c_{i_\alpha}(\Omega)]^\dagger \rho_S^I]e^{i(\Omega-\Omega^\prime)t}\right.\\+\left.(H^\alpha(\Omega)-iH^{\alpha I}(\Omega))\times[[c_{i_\alpha}(\Omega^\prime)]^\dagger,c_{i_\alpha}(\Omega)\rho_S^I]e^{-i(\Omega-\Omega^\prime)t}+\text{H.c.} \right\rbrace. \label{AppRedfield}
\end{multline}
\end{widetext}
Transforming back to Schr\"odinger picture one obtains Eq. (\ref{FullRedfield1}) and Eq. (\ref{FullRedfield2}).

\section{Generalization of the Redfield equation to three dimensions}\label{AppRed3D}

For a three-dimensional system, we generalize the system-bath Hamiltonian to:
\begin{equation}
H_{SB}^{(\alpha)}=\sum_{\lambda,\mathbf{R_\alpha}}(T_{\lambda,\mathbf{R_\alpha}}^\alpha f_{\lambda}^{\alpha \dagger} c_{\mathbf{R_\alpha}}+\text{H.c.}), \quad \alpha=L\  \text{or}\  R
\end{equation}
where $\mathbf{R_\alpha}$ in general denotes the coordinate, and other degrees of freedom (such as orbital) of the state connected to bath $\alpha$. Similar calculations yield a formula analogous to Eq. (\ref{drhodt}): 
\begin{multline}
\frac{d\rho^I_S}{dt}=-\sum_{\alpha,\mathbf{R_\alpha},\mathbf{R_\alpha^\prime}}\int_{0}^t dt^\prime\left( [c_{\mathbf{R_\alpha}}(t),\hat{C}^{\alpha+}(t,t^\prime;\mathbf{R_\alpha},\mathbf{R_\alpha^\prime})\rho^I_S(t^\prime)]\right. \\+\left. [c_{\mathbf{R_\alpha}}^\dagger(t),\hat{C}^{\alpha-}(t,t^\prime;\mathbf{R_\alpha},\mathbf{R_\alpha^\prime})\rho^I_S(t^\prime)]+\text{H.c.}\right)
\end{multline}
where
\begin{equation}
\hat{C}^{\alpha+}(t,t^\prime;\mathbf{R_\alpha},\mathbf{R_\alpha^\prime})=\sum_\lambda  T^\alpha_{\lambda,\mathbf{R_\alpha}}T^{\alpha*}_{\lambda,\mathbf{R_\alpha^\prime}}\langle f_\lambda^{\alpha\dagger}(t)f^\alpha_\lambda(t^\prime)\rangle_B c_{\mathbf{R_\alpha^\prime}}^\dagger(t^\prime), 
\end{equation}
\begin{equation}
\hat{C}^{\alpha-}(t,t^\prime;\mathbf{R_\alpha},\mathbf{R_\alpha^\prime})=\sum_\lambda  T_{\lambda,\mathbf{R_\alpha}}^{\alpha*}T^{\alpha}_{\lambda,\mathbf{R_\alpha^\prime}}\langle f_\lambda^{\alpha}(t)f^{\alpha\dagger}_\lambda(t^\prime)\rangle_B c_{\mathbf{R_\alpha^\prime}}(t^\prime).
\end{equation}
We could define the generalized spectral density
\begin{equation}
 N^\alpha(\mathbf{R_\alpha},\mathbf{R_\alpha^{\prime}};\omega)=\sum_{\lambda}T_{\lambda,\mathbf{R_\alpha}}T_{\lambda,\mathbf{R_\alpha^\prime}}^*\delta(\omega-\epsilon^\alpha_\lambda)
\end{equation}
and another calculation that includes a transformation back to Schr\"odinger picture yields the three-dimensional Redfield equation in real space:
\begin{widetext}
\begin{multline}
\frac{d\rho_S^S}{dt}=-i[H_S,\rho_S^S]-\sum_{\alpha,\mathbf{R_\alpha},\mathbf{R_\alpha^\prime},\Omega}\left\lbrace (F^\alpha(\mathbf{R_\alpha},\mathbf{R_\alpha^\prime},\Omega)+iF^{\alpha I}(\mathbf{R_\alpha},\mathbf{R_\alpha^\prime},\Omega))\times[c_{\mathbf{R_\alpha}},[c_{\mathbf{R_\alpha^\prime}}(\Omega)]^\dagger \rho_S^S]\right.\\+\left.(H^\alpha(\mathbf{R_\alpha^\prime},\mathbf{R_\alpha},\Omega)-iH^{\alpha I}(\mathbf{R_\alpha^\prime},\mathbf{R_\alpha},\Omega))\times[c_{\mathbf{R_\alpha}}^\dagger,c_{\mathbf{R_\alpha^\prime}}(\Omega)\rho_S^S]+\text{H.c.} \right\rbrace \label{3dRedfield}
\end{multline}
\end{widetext}
where for example
\begin{equation}
F^\alpha(\mathbf{R_\alpha},\mathbf{R_\alpha^\prime},\Omega)=\pi N^\alpha(\mathbf{R_\alpha},\mathbf{R_\alpha^\prime},\Omega)f^\alpha(\Omega),
\end{equation}
\begin{equation}
F^{\alpha I}(\mathbf{R_\alpha},\mathbf{R_\alpha^\prime},\Omega)=\frac{1}{\pi}\text{P}\int_{-\infty}^{\infty}d\omega\frac{F^\alpha(\mathbf{R_\alpha},\mathbf{R_\alpha^\prime},\omega)}{\omega-\Omega}
\end{equation}
and a similar definition holds for $H^\alpha(\mathbf{R_\alpha},\mathbf{R_\alpha^\prime},\Omega)$ and $H^{\alpha I}(\mathbf{R_\alpha},\mathbf{R_\alpha^\prime},\Omega)$. To simplify the above equation, we assume that the whole model, including both baths and system, has translational invariance along the transverse direction and there is no interaction. We separate $\mathbf{R_\alpha}$ into two parts: $\mathbf{R_\alpha^\perp}$ which represents the coordinate of the unit cell in the transverse direction, and $R_\alpha^\parallel$ along the junction. Due to the translational invariance in transverse direction, all functions depend only on the difference $\mathbf{R_\alpha^\perp}-\mathbf{R_\alpha^{\perp\prime}}$ and we can define the Fourier transform for a general function $g(\mathbf{x_\perp};x_\parallel)=g(\mathbf{R_\alpha^\perp}-\mathbf{R_\alpha^{\perp\prime}};x_\parallel)$
\begin{equation}
g^\alpha(\mathbf{k_\perp};x_\parallel)=\sum_{\mathbf{x_\perp}} e^{i\mathbf{k_\perp} \cdot \mathbf{x_\perp}} g^\alpha(\mathbf{x_\perp};x_\parallel)
\end{equation}
and its inverse
\begin{equation}
g^\alpha(\mathbf{x_\perp};x_\parallel)=\sum_{\mathbf{k_\perp}}\frac{e^{-i\mathbf{k_\perp} \cdot \mathbf{x_\perp}}g^\alpha(\mathbf{k_\perp};x_\parallel)}{N_xN_y}
\end{equation}\
where $N_x$ and $N_y$ are the number of unit cells in $x$ and $y$ direction respectively. Translational symmetry also means that the energy eigenstates can be labelled by the transverse momentum $\mathbf{k_\perp}$ and the longitudinal index $\lambda_{\mathbf{k_\perp}}$. Therefore, we have
\begin{equation}
c_{\mathbf{R_\alpha^\prime}}^\dagger(\Omega)=\sum_{\epsilon(\mathbf{k_\perp},\lambda_{\mathbf{k_\perp}})=\Omega}\frac{e^{-i\mathbf{k_\perp} \cdot\mathbf{R_\alpha^{\perp\prime}}}}{\sqrt{N_xN_y}}\langle \lambda_{\mathbf{k_\perp}}\vert R_\alpha^{\parallel\prime}\rangle c_{\mathbf{k_\perp},\lambda_{\mathbf{k_\perp}}}^\dagger
\end{equation}
and
\begin{equation}
c_{\mathbf{R_\alpha}}=\sum_{\mathbf{k_\perp}}\frac{e^{i\mathbf{k_\perp} \cdot\mathbf{R_\alpha^\perp}}}{\sqrt{N_xN_y}} c_{\mathbf{k_\perp},R_\alpha^\parallel}
\end{equation}
Substituting these equations into Eq. (\ref{3dRedfield}) and summing over $\mathbf{R^\perp_\alpha}$ and $\mathbf{R^{\perp\prime}_\alpha}$ yields
\begin{widetext}
\begin{multline}
\frac{d\rho_S}{dt}=-i[H_S,\rho_S^S]-\sum_{\alpha}\sum_{\mathbf{k_\perp},\lambda_{\mathbf{k_\perp}}}\sum_{R_\alpha^\parallel,R_\alpha^{\parallel\prime}}\left\lbrace (F^\alpha(\mathbf{k_\perp};R_\alpha^\parallel,R_\alpha^{\parallel\prime},\Omega)+iF^{\alpha I}(\mathbf{k_\perp};R_\alpha^\parallel,R_\alpha^{\parallel\prime},\Omega))\times[c_{\mathbf{k_\perp},R_\alpha^\parallel},c_{\mathbf{k_\perp},\lambda_{\mathbf{k_\perp}}}^\dagger \rho_S]\times \langle \lambda_{\mathbf{k_\perp}}\vert R_\alpha^{\parallel\prime}\rangle\right.\\+\left.(H^\alpha(\mathbf{k_\perp};R_\alpha^{\parallel\prime},R_\alpha^{\parallel},\Omega)-iH^{\alpha I}(\mathbf{k_\perp};R_\alpha^{\parallel\prime},R_\alpha^{\parallel},\Omega))\times[c_{\mathbf{k_\perp},R_\alpha^\parallel}^\dagger,c_{\mathbf{k_\perp},\lambda_{\mathbf{k_\perp}}}\rho_S]\times \langle R_\alpha^{\parallel\prime}\vert\lambda_{\mathbf{k_\perp}} \rangle+\text{H.c.} \right\rbrace 
\end{multline}
\end{widetext}
where $\Omega=\epsilon(\mathbf{k_\perp},\lambda_{\mathbf{k_\perp}})$ is implied. We may also obtain the EOM for the OPDM. Ignoring the dependence of $N^\alpha(\mathbf{k_\perp};R_\alpha^\parallel,R_\alpha^{\parallel\prime},\Omega)$ on $\Omega$ and neglecting all CPVPs, following Section \ref{SecModified} we obtain
\begin{multline}
\frac{d\rho^{(1)}}{dt}=-i[h,\rho^{(1)}]+\sum_{\alpha,\mathbf{k_\perp}}[(\rho_{\mathbf{k_\perp},\alpha}^{(1)}-\rho^{(1)})P_{\mathbf{k_\perp},\alpha}+\text{H.c.}]
\end{multline}
where 
\begin{equation}
\rho_{\mathbf{k_\perp},\alpha}^{(1)}=\sum_{\lambda_{\mathbf{k_\perp}}}\vert \mathbf{k_\perp},\lambda_{\mathbf{k_\perp}}\rangle f^\alpha(\epsilon(\mathbf{k_\perp},\lambda_{\mathbf{k_\perp}}))\langle \mathbf{k_\perp},\lambda_{\mathbf{k_\perp}}\vert
\end{equation}
and
\begin{equation}
P_{\mathbf{k_\perp},\alpha}=\sum_{R_\alpha^\parallel,R_\alpha^{\parallel\prime}}\pi N^\alpha(\mathbf{k_\perp},R_\alpha^\parallel,R_\alpha^{\parallel\prime})\vert \mathbf{k_\perp},R_\alpha^{\parallel\prime}\rangle\langle \mathbf{k_\perp},R_\alpha^\parallel\vert.
\end{equation} 

\section{Generalization of NEGF to three dimensions}\label{AppGreen3D}

With the same assumption and notations as in Appendix \ref{AppRed3D} the Green's functions describing the system can now be expressed as $G_{R^\parallel,R^{\parallel \prime}}({\bf k}_{\perp},\omega)$ due to translational 
invariance in $x-y$ direction. The Dyson equations for the Keldysh Green's functions of the chain coupled to the baths now read:
\begin{eqnarray}
G^{r,a}({\bf k}_{\perp},\omega)=(\omega \pm i \eta -h({\bf k}_{\perp}) -\Sigma^{r,a}(\omega))^{-1},
\nonumber \\
G^<({\bf k}_{\perp},\omega)=G^r({\bf k}_{\perp},\omega) \Sigma^<(\omega) G^a({\bf k}_{\perp},\omega),
\end{eqnarray}
where the self-energies of the chain due to the coupling to the baths are given by
\begin{eqnarray}
\Sigma_{R^\parallel,R^{\parallel\prime}}^{r,a}(\omega)= \mp i \sum_{\alpha={L,R}}\sum_{R_\alpha^\parallel} J^\alpha  \delta_{R^\parallel,R_\alpha^\parallel} \delta_{R^{\parallel\prime},R_\alpha^\parallel}
\nonumber \\
\Sigma_{R^\parallel,R^{\parallel\prime}}^<(\omega)=2i \sum_{\alpha={L,R}}\sum_{R_\alpha^\parallel}J^\alpha f^\alpha(\omega) \delta_{R^\parallel,R_\alpha^\parallel} \delta_{R^{\parallel\prime},R_\alpha^\parallel}, 
\label{eq:selfener3d}
\end{eqnarray}
where we have again made the wide-band approximation for the baths.

In the three-dimensional model, the occupations along the junction read:
\begin{eqnarray}
\langle n_{R^\parallel} \rangle &=& -{i \over N_x N_y} \sum_{{\bf k}_{\perp}}\int_{-\infty}^{\infty} {d \omega \over 2 \pi} G^<_{R^\parallel,R^\parallel}({\bf k}_{\perp},\omega),
\end{eqnarray}
The occupations are again split into equilibrium and non-equilibrium contributions. The equilibrium part reads:
\begin{eqnarray}
\langle n^{\alpha,eq}_{R^\parallel} \rangle &=& -{1 \over N_x N_y} \sum_{{\bf k}_{\perp}} \int_{-\infty}^{\infty} {d \omega \over \pi} \text{Im} G^r_{R^\parallel,R^\parallel}({\bf k}_{\perp},\omega) f^\alpha(\omega)
\nonumber \\
&=& {1 \over \beta} {1 \over N_x N_y} \sum_{{\bf k}_{\perp}, n} e^{i \omega_n 0^+} G^{\mu_\alpha}_{R^\parallel,R^\parallel}({\bf k}_{\perp},i\omega_n),
\end{eqnarray}
while the non-equilibrium contribution is:
\begin{widetext}
\begin{eqnarray}
\langle n^{R,neq}_{R^\parallel,\sigma} \rangle = {2J^R  \over N_x N_y} \sum_{{\bf k}_{\perp},R^\parallel_R} \int_{-\infty}^{\infty} {d \omega \over 2 \pi} (f^R(\omega) -f^L(\omega)) |G^r_{R^\parallel,R^\parallel_R}({\bf k}_{\perp},\omega)|^2,
 \nonumber \\
 \langle n^{L,neq}_{R^\parallel,\sigma} \rangle = {2J^L \over N_x N_y} \sum_{{\bf k}_{\perp},R^\parallel_L} \int_{-\infty}^{\infty} {d \omega \over 2 \pi} (f_L(\omega) -f_R(\omega)) |G^r_{R^\parallel,R^\parallel_L}({\bf k}_{\perp},\omega)|^2.
\end{eqnarray}  
\end{widetext}

Finally, the expression for the particle current through a unit cell becomes:
\begin{widetext}
\begin{equation}
j=2 J^L J^R {1 \over N_x N_y}\sum_{{\bf k}_{\perp}} \sum_{R^\parallel_L,R^\parallel_R} \int_{-\infty}^{\infty} {d \omega \over 2 \pi} \left(f_L(\omega)-f_R(\omega)\right) |G^r_{R^\parallel_L,R^\parallel_R}({\bf k}_{\perp},\omega)|^2.
\label{eq:current}
\end{equation}  
\end{widetext}

\section{Perturbative solution to the Redfield equation for the OPDM} \label{AppPerturbation}
Consider a system that may have some degeneracies and obeys Eq. (\ref{ODDMRedfield2}). If the system-bath coupling $J$ is small, or in other words $J\ll \epsilon_\alpha -\epsilon_\beta$ for any $\alpha \neq\beta$ that $\epsilon_\alpha\neq \epsilon_\beta$, we can expand $\rho=\rho^0+J\rho^1+J^2\rho^2+...$ and solve the equation order by order. To zeroth order, it is easy to show that the OPDM $\rho^0$ is block diagonalized in the energy eigenbasis and each block corresponds to a degenerate space. To first order, 
\begin{equation}
0=-i[h,\rho^1]+\sum_\alpha\{\rho_\alpha,P_{i_\alpha}\}-\{\rho^0,P\}
\end{equation}
where we have used the shorthand $P=\sum_\alpha P_{i_\alpha}$. We may also write the above equation in block matrix form:
\begin{equation}
\rho^0_{uu}P_{uu}+P_{uu}\rho^0_{uu}=2\sum_\alpha (\rho_\alpha)_{uu}(P_{i_\alpha})_{uu} \label{B2}
\end{equation}
and
\begin{widetext}
\begin{equation}
\rho^1_{uw}=\frac{\sum_\alpha[(\rho_\alpha)_{uu}(P_{i_\alpha})_{uw}+(P_{i_\alpha})_{uw}(\rho_\alpha)_{ww}-\rho^0_{uu}(P_{i_\alpha})_{uw}-(P_{i_\alpha})_{uw}\rho^0_{ww}]}{i(\epsilon_u-\epsilon_w)}, \quad\text{for} \  u\neq w \label{B3}
\end{equation}
\end{widetext}
where indices $u$ and $w$ label the blocks corresponding to different degenerate eigenspaces.  Working in the eigenbasis that diagonalizes $P$ in each block, as long as $P$ does not have opposite eigenvalues, one can solve for $\rho^0$ in each degenerate eigenspace leading to:
\begin{equation}
\rho^0_{ij}=2\frac{\sum_\alpha \rho_\alpha (P_{i_\alpha})_{ij}}{P_{jj}+P_{ii}}. \label{B5}
\end{equation}
In particular, if there is no degeneracy, Eq. (\ref{B5}) becomes Eq. (\ref{DiagPert}).

\end{document}